\documentclass[aps,prd,amssymb,twocolumn,superscriptaddress,floatfix,nofootinbib,10pt]{revtex4}
\usepackage{graphicx}
\usepackage{subfigure}
\usepackage{hyperref}
\usepackage{amsmath}
\usepackage{physics}
\usepackage{hyphenat}
\usepackage{amsfonts}
\usepackage{booktabs}
\usepackage{amssymb}
\usepackage[usenames]{color}
\usepackage{lineno}
\usepackage{bm}
\usepackage{mathptmx}
\usepackage[normalem]{ulem}
\usepackage{orcidlink}

\usepackage{graphics}

\newcommand{\figcaption}{\def\@captype{figure}\caption}
\newcommand{\tabcaption}{\def\@captype{table}\caption}

\begin{document}

\title{On the possible existence of a $S=-3, \, I=1$ pentaquark}

\author{Albert Feijoo\orcidlink{0000-0002-8580-802X}}
 \thanks{Corresponding author}
	\email{edfeijoo@ific.uv.es}
	\affiliation{Physik Department E62, Technische Universit\"at M\"unchen, Garching, Germany, EU}
    \affiliation{Instituto de F\'{i}sica Corpuscular, Centro Mixto Universidad de Valencia-CSIC, Institutos de Investigaci\'{o}n de Paterna, Aptdo. 22085, E-46071 Valencia, Spain}

\author{Isaac Vida\~na\orcidlink{0000-0001-7930-9112}}
\email{isaac.vidana@ct.infn.it}
	\affiliation{Istituto Nazionale di Fisica Nucleare, Sezione di Catania, Dipartimento di Fisica ``Ettore Majorana'', Universit\`a di Catania, Via Santa Sofia 64, I-95123 Catania, Italy}

\begin{abstract}
We analyze the possible existence of a strangeness $S=-3$, isospin $I=1$ pentaquark  state $P_{sss}$ generated dynamically from the $\bar{K}\Xi$ interaction. We employ a unitarized scheme in coupled channels based on the chiral Lagrangian expanded up to next-to-leading order (NLO), and show that the inclusion of the NLO terms is crucial to provide the necessary attraction that favors the existence of such triply strange pentaquark. The $\bar{K}\Xi$ femtoscopic correlation functions are calculated as example of a possible experimental measurement in which a direct signal of the $P_{sss}$ state could be observed.

\end{abstract}

\date{\today}

\maketitle

\textit{Introduction:} Over the last decade, the hadron spectroscopy has experienced the striking irruption of the experimental evidences for pentaquark structures, which has turned into one of the major discoveries in this field. The success of this quest is largely due to the role played by the LHCb collaboration which in 2015 reported the first 
two $J/\psi p$ resonant structures seen in the corresponding invariant mass associated with the $\Lambda_b \to J/\psi K^- p$ decay~\cite{LHCb:2015yax}. These states, lately named $P_{\psi}^N(4380)$ and $P_{\psi}^N(4450)$, cannot be interpreted as excited nucleon resonances given their large masses, which makes mandatory the presence of a $\bar{c}c$ pair in their inner quark structure. Actually, these hidden charm pentaquarks were already predicted either within the molecular picture~\cite{Wu:2010jy,Wu:2010vk,Yang:2011wz,Xiao:2013yca,Karliner:2015ina} or employing a quark model approach~\cite{Wang:2011rga,Yuan:2012wz}. An ulterior LHCb analysis combining Run~I and Run~II data found a new pentaquark and disclosed a double-peak structure masked in the original $P_{\psi}^N(4450)$ signal thereby leaving four more accurate hidden charm pentaquarks: $P_{\psi}^N(4312)$, $P_{\psi}^N(4380)$, $P_{\psi}^N(4440)$ and $P_{\psi}^N(4457)$. Slightly after, again the LHCb collaboration provided the evidence for an extra $P_{\psi}^N(4337)$ structure present in the $J/\psi p$ and $J/\psi \bar{p}$ invariant mass distributions of the $B^0_s \to J/\psi p\bar{p}$ decay \cite{LHCb:2021chn}. All these observations triggered a lot of activity in the theoretical community which precipitated in a plethora of studies based on different interpretations: as molecular states \cite{Chen:2015loa,He:2015cea,Liu:2019tjn,Du:2021fmf}, using different quark configurations \cite{Chen:2015moa,Wang:2015epa,Wang:2019got,Ortega:2016syt,Park:2017jbn,Weng:2019ynv,Zhu:2019iwm,Deng:2022vkv}, as triangle singularities \cite{Guo:2015umn,Liu:2015fea,Mikhasenko:2015vca}, or as cusp-like structures \cite{Nakamura:2021dix}.

After these experimental breakthroughs, it seemed unlikely that there would be no strangeness $S=-1$ partners of such states. As a matter of fact, by that time there were already a vast spectra of such molecular states predicted employing either $SU(4)$ symmetry and extended hidden gauge unitarized approaches \cite{Hofmann:2005sw,Wu:2010vk,Wang:2019nvm} or alternatively predicted with diquark--diquark--antiquark models \cite{Anisovich:2015zqa,Wang:2015wsa}. Moreover, in Refs.~\cite{Feijoo:2015kts,Lu:2016roh,Chen:2015sxa,Shen:2020gpw}, the exploration of the $J/\psi \Lambda$ invariant mass distribution associated to some $\Lambda_b$ and $\Xi_b$ decay processes were suggested as suitable reactions to find hidden charm strange states. Finally, the confirmation of the $P_{\psi s}^N$ states arrived with the experimental analysis of the $J/\psi \Lambda$ invariant mass from the $\Xi_b^-\to J/\psi \Lambda K^-$ \cite{LHCb:2020jpq} and $B^-\to J/\psi \Lambda \bar{p}$ \cite{LHCb:2022ogu} decays. With the observation of the $P_{\psi s}^N(4459)$ and $P_{\psi s}^N(4338)$ pentaquarks, their nature has become a hot topic that has boosted reviews of the existent theoretical machinery or motivated novel approaches \cite{Chen:2020uif,Chen:2020opr,Liu:2020hcv,Feijoo:2022rxf,Karliner:2022erb,Wang:2022mxy,Yan:2022wuz,Ozdem:2022kei,Ortega:2022uyu,Wang:2022tib,Peng:2020hql,Xiao:2021rgp,Zhu:2021lhd,Burns:2022uha,Nakamura:2022gtu,Meng:2022wgl,Wu:2024lud}.
Within this promising context, the natural next step is to turn the attention to the hidden-charm pentaquark with double strangeness, whose experimental evidence is still pending despite its theoretical predictions \cite{Wang:2020bjt,Ferretti:2020ewe,Azizi:2021pbh,Ortega:2022uyu,Marse-Valera:2022khy,Roca:2024nsi,Song:2024yli}. 

Being chronologically accurate, the confirmation of the pentaquark conjecture was claimed for the first time with the peak signal appearing at $1.54$~GeV in the $K^+n$ invariant mass distribution from the $\gamma n \to K^+K^-n$ process on $^{12}C$ \cite{LEPS:2003wug}. This structure, known as $\Theta^+(1540)$ pentaquark and with minimal quark content $\ket{uudd\bar{s}}$, was explained in Refs.~\cite{MartinezTorres:2010zzb,MartinezTorres:2010xqq} as an artificial broad peak due to the method used in the experiment to associate momenta to the undetected proton and neutron together with the chosen cuts. Furthermore, several experimental groups found no evidence of the $\Theta^+(1540)$ performing experiments where it was expected to be seen. Nevertheless, this fact does not rule out the possibility of finding similar structures in other light sectors. 

The $S=-3$ hadron spectroscopy has been barely explored so far.  A clear proof of the lack of knowledge on this hadron spectrum is the presence of only four $\Omega^*$ states, with $2$- or $3$-star status and with no assigned spin-parity quantum numbers, in the PDG compilation \cite{ParticleDataGroup:2024cfk}. On the theoretical side, the dynamical generation of $S=-3$ baryon resonances from unitarized effective field theories in coupled channels (CC) was studied in \cite{Kolomeitsev:2003kt,Xu:2015bpl} employing pseudoscalar mesons and ground state baryons of the decuplet ($J^P=3/2^+$), while the $\bar{K}\Xi$ interaction was investigated within an extended chiral $SU(3)$ quark model by solving a resonating group method equation \cite{Wang:2008zzz}. However, in our opinion, the most relevant study in this theoretical context was the one presented in Ref.~\cite{Gamermann:2011mq}, where a consistent $SU(6)$ extension of the meson-baryon chiral Lagrangian within a CC unitary approach is employed. There, a wider spectrum of meson-baryon channels was considered combining all possible pairs formed by pseudoscalar and vector mesons with baryons of the ground state octet and the decuplet.
The authors predicted nine $\Omega^*$ resonances, two of them compatible with the observed $\Omega(2250)$ and $\Omega(2380)$. This work included also a prediction of the number of exotic states with $S=-3$ expected to be generated on the basis of the attractive multiplets of the model. This number amounts to seven states, three of which have $J^P=1/2^-$.

In this letter, we present a unitarized scheme in CC, only accounting for pseudoscalar mesons and the ground state baryons of the octet, whose interaction is purely based on the chiral Lagrangian expanded up NLO. As it is shown below, the inclusion of higher order corrections, never taken into account before in the $S=-3$ sector, plays a crucial role in the dynamical generation of a triply strange pentaquark, $P_{sss}$, and a molecular $\Omega^*$ state. Actually, the leading order (LO) contributions provide a very shallow repulsive interaction, while the NLO terms confer the necessary attractive character that depending on the values of NLO driving parameters, the so-called low energy constants (LECs), can favor the presence of such poles in the scattering amplitudes. \\

\textit{Formalism and Discussion:} Unitarized chiral perturbation theory (UChPT) has shown to be a very suitable tool to address the hadron-hadron scattering at energies around resonances and bound states. This nonperturbative scheme guarantees the convergence, as well as the unitarity and analyticity of the scattering amplitude. In the present work, we implement unitarity by solving the 
Bethe--Salpeter~(BS) equation with CC. As it is shown in~\cite{Oset:1997it,Hyodo:2011ur}, the interaction kernel can be conveniently split into its on-shell contribution and the corresponding off-shell one. The off-shell part gives rise to a tadpole-type diagram which can be reabsorbed into renormalization of couplings and masses and, hence, be omitted from the calculation. This method allows one to factorize the interaction kernel and the scattering amplitude out of the integral equation, transforming a complex system of coupled integral equations into a simple system of algebraic equations whose matrix form reads:
\begin{equation}
T_{ij} ={(1-V_{il}G_l)}^{-1}V_{lj} ,
 \label{T_algebraic}
\end{equation}
where $V_{ij}$ is the driving interaction kernel, $T_{ij}$ is the related scattering amplitude for the transition from an $i$-th channel to a $j$-th one, while $G_l$ stands for the loop function of the propagating channel $l$. As this function diverges logarithmically, a dimensional regularization scheme is applied which leads to:
\begin{eqnarray}
 G_l & = &\frac{2M_l}{(4\pi)^2} \Bigg \lbrace a_l(\mu)+\ln\frac{M_l^2}{\mu^2}+\frac{m_l^2-M_l^2+s}{2s}\ln\frac{m_l^2}{M_l^2}  \nonumber \\ 
 &  +   &  \frac{q_{\rm cm}}{\sqrt{s}}\ln\left[\frac{(s+2\sqrt{s}q_{\rm cm})^2-(M_l^2-m_l^2)^2}{(s-2\sqrt{s}q_{\rm cm})^2-(M_l^2-m_l^2)^2}\right]\Bigg \rbrace,  
 \label{dim_reg}    
\end{eqnarray}
where $M_l$ and $m_l$ are, in the present case, the baryon and meson masses of channel $l$, and $q_{cm}$ is the center-of-mass (CM) momentum of the channel pair at an energy $\sqrt{s}$. The subtraction constants (SCs) $a_l(\mu)$ replace the divergence for a given dimensional regularization scale $\mu$. There are as many SCs as channels considered in the sector at hand, but this number can be reduced applying isospin symmetry arguments. The pseudoscalar meson-ground state baryons channels in the $S=-3$ sector are shown in Tab.\ \ref{tab:channels}. Since in this sector there is no available data, following Ref.~\cite{Oller:2000fj}), we assign a natural size value for the SCs and fix all of them to be approximately $-2$ when taking $\mu=630$~MeV.

\begin{table}[t]
\centering
{
\begin{tabular}{c|c|c}
\hline
\hline 
$Q=0$ & $Q=-1$ & $Q=-2$ \\
\hline
$\bar{K}^0\Xi^0(1813)$ &
\,$K^-\Xi^0(1809)\, \, \, \, \bar{K}^0\Xi^-(1819)$\, &
$K^-\Xi^-(1815)$  \\
\hline
\hline
\end{tabular}}
\caption{Pseudoscalar meson-ground state baryon channels in the strangeness $S=-3$ sector. The value of the energy threshold of each channel is indicated in brakets.}
\label{tab:channels}
\end{table}

\begin{figure*}[t]
\begin{center}
\includegraphics[width=0.75\textwidth,keepaspectratio]{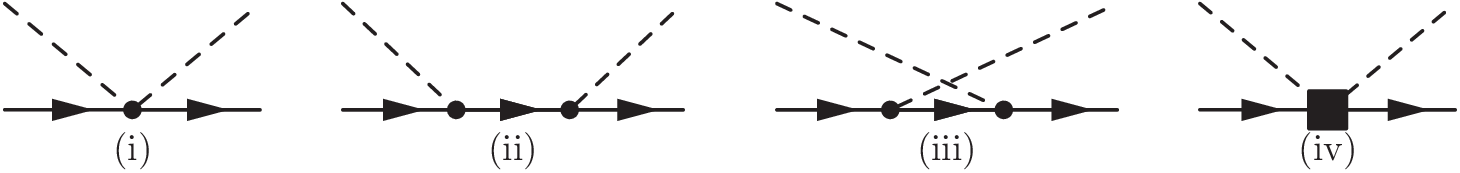}
\caption{Feynman diagrams for meson-baryon interaction: WT term (i), direct and crossed Born terms (ii) and (iii), and NLO terms (iv). Dashed (solid) lines represent the pseudoscalar octet mesons (octet baryons). }
\label{fig:diagrams}
\end{center}
\end{figure*}

The pseudoscalar meson-ground state baryon kernel is derived from the $SU(3)$ effective chiral Lagrangian, which provides the fundamental blocks of the interaction that preserve the symmetries of QCD employing hadron fields as degrees of freedom. These constituent pieces are arranged in an expansion by order of relevance following a power counting scheme (see~\cite{Scherer:2002tk} for a more detailed explanation). The most general expression for the $SU(3)$ chiral effective Lagrangian up to NLO can be written as $\mathcal{L}_{\phi B}^{eff}=\mathcal{L}_{\phi B}^{(1)}+\mathcal{L}_{\phi B}^{(2)}$, with $\mathcal{L}_{\phi B}^{(1)}$ and $\mathcal{L}_{\phi B}^{(2)}$ accounting for the LO contributions and the $\cal{O}$$(p^2)$ corrections in the expansion, respectively. Both pieces can be expressed as follows:  
\begin{eqnarray} 
\mathcal{L}_{\phi B}^{(1)} & = & i \langle \bar{B} \gamma_{\mu} [D^{\mu},B] \rangle
                            - M_0 \langle \bar{B}B \rangle  
                           + \frac{1}{2} D \langle \bar{B} \gamma_{\mu} 
                             \gamma_5 \{u^{\mu},B\} \rangle \nonumber \\
                  & &      + \frac{1}{2} F \langle \bar{B} \gamma_{\mu} 
                               \gamma_5 [u^{\mu},B] \rangle \ ,
\label{LagrphiB1} \\
  \mathcal{L}_{\phi B}^{(2)}& = & b_D \langle \bar{B} \{\chi_+,B\} \rangle
                             + b_F \langle \bar{B} [\chi_+,B] \rangle
                             + b_0 \langle \bar{B} B \rangle \langle \chi_+ \rangle  \nonumber \\ 
                     &  & + d_1 \langle \bar{B} \{u_{\mu},[u^{\mu},B]\} \rangle 
                            + d_2 \langle \bar{B} [u_{\mu},[u^{\mu},B]] \rangle     \nonumber \\
                    &  &  + d_3 \langle \bar{B} u_{\mu} \rangle \langle u^{\mu} B \rangle
                            + d_4 \langle \bar{B} B \rangle \langle u^{\mu} u_{\mu} \rangle \ .
\label{LagrphiB2}
\end{eqnarray}

As main inputs entering in Eqs.~(\ref{LagrphiB1}) and (\ref{LagrphiB2}), $B$ stands for the octet baryon matrix, while the matrix of the pseudoscalar mesons $\phi$ is implicitly contained in $u_\mu = i u^\dagger \partial_\mu U u^\dagger$, where $U(\phi) = u^2(\phi) = \exp{\left( \sqrt{2} {\rm i} \phi/f \right)} $ with $f$ being the effective meson decay constant. The covariant derivative is given by  $[D_\mu, B] = \partial_\mu B + [ \Gamma_\mu, B]$, with the connection given by $\Gamma_\mu =  [ u^\dagger,  \partial_\mu u] /2$. From this last piece, the so-called Weinberg--Tomozawa (WT) contribution is computed, whose schematic representation is shown by the contact diagram (i) of Fig.~\ref{fig:diagrams}. $M_0$ stands for the baryon octet mass in the chiral limit. However, here we employ the physical hadron masses
as it is usually done in this kind of approaches. The last two terms
in Eq.~(\ref{LagrphiB1}), containing the pseudoscalar coupling and preceded by the axial vector constants $D$ and $F$, are needed to calculate the contribution of the direct (s-channel) and crossed (u-channel) Born terms represented, respectively, by the diagrams (ii) and (iii).
On the other hand, the tree-level NLO contributions are fully extracted from $\mathcal{L}_{\phi B}^{(2)}$ and are schematically depicted in diagram (iv). The first three terms in Eq.\ (\ref{LagrphiB2}) break explicitly chiral symmetry, being $\chi_+ = 2 B_0 (u^\dagger \mathcal{M} u^\dagger + u \mathcal{M})$ with $\mathcal{M} = {\rm diag}(m_u, m_d, m_s)$) and $B_0=-\langle 0|\bar{q}q|0\rangle /f^2$.
The coefficients $b_D$, $b_F$, $b_0$ and $d_i$ $(i=1,\dots,4)$  preceding the NLO terms are LECs that are not established by the symmetries of the underlying theory, thereby leaving them as parameters of the model to be determined from experimental measurements. It is pertinent to comment that the parameters accompanying the terms proportional to the $\chi_+$ field should fulfill constraints related to the mass splitting of baryons. Complementarily, the coefficient $b_0$ could be extracted either from the pion-nucleon sigma term or from the strangeness content of the proton \cite{Gasser:1990ce}. Nevertheless, since our study goes beyond tree-level because of the implementation of the coupled-channel unitarization and the explicit use of the physical hadron masses in the calculations, it is hard to think about a possible universality of the NLO LECs values. But, given the lack of scattering data in the $S=-3$ sector, one has no possibility to constrain the LECs properly. Therefore, as done in the study presented in \cite{Feijoo:2023wua} applied to the $S=-2$ sector, the $SU(3)$ symmetry is effectively assumed to assign the values of $f$, $D$, $F$ and the LECs $b_D$, $b_F$, $b_0$ and $d_{1 ... 4}$.
In particular, we take these values from the BCN \cite{Feijoo:2018den} and VBC \cite{Sarti:2023wlg} models
to have larger casuistry. In the BCN model, the parameters were obtained fitting the model to a large set of low-energy scattering data from $K^-p\to \phi B$ ($S=-1$) processes, as well as, the $\bar{K}N$ threshold observables typically employed in these studies (branching ratios and the scattering length extracted from the very precise SIDDHARTA measurements \cite{SIDD}). On the other hand, the characterizing parametrization of the VBC model was extracted from the high-precision $K^-\Lambda$ femtoscopic data \cite{ALICE:2023wjz}. The study of Ref.~\cite{Sarti:2023wlg} has been the only attempt to determine the NLO LECs in the $S=-2$ sector so far.

The total interaction kernel derived from Eqs.~(\ref{LagrphiB1}) and (\ref{LagrphiB2}) can be expressed as $V_{ij}=V^{\scriptscriptstyle WT}_{ij}+V^{D}_{ij}+V^{C}_{ij}+V^{\scriptscriptstyle NLO}_{ij}$, where the analytical form 
of each contribution
can be found {\it e.g.,} in Refs.~\cite{Borasoy:2005ie,Hyodo:2011ur,Ramos:2016odk,Feijoo:2021zau}. 
In particular, here we employ the expressions given in Eqs.~(7)--(10) of Ref.\ \cite{Feijoo:2021zau} that, once projected onto $s-$wave, reduce to those of Eqs.~(6)--(8) and (10) in Ref.\ \cite{Ramos:2016odk}. 
The expressions of
$V^{\scriptscriptstyle WT}_{ij},V^{D}_{ij},V^{C}_{ij}$ and 
$V^{\scriptscriptstyle NLO}_{ij}$ are preceded by 
coefficients which encode the symmetries of the interaction and should be calculated for each charge channel of the $S=-3$ sector considered here.  
The $C_{ij}$ coefficients of the WT contact potential, as well as, the NLO Clebsch--Gordan type coefficients, $D_{ij}$ and $L_{ij}$, are displayed in Tables~\ref{coeff_ Qm1} and \ref{coeff_ Qm2_Q0}. The vertices involving the baryon-meson-baryon triads in both Born 
terms
($V^{D}_{ij}$ direct and $V^{C}_{ij}$ crossed) can be computed taking into account the relations (A.5) in Appendix~A of Ref.~\cite{Borasoy:2005ie}, which come as linear combinations of the axial vectors constants. It is convenient to mention here that there is no contribution of the direct Born term (s-channel diagram) since no ground state baryon with $J^P=1/2^-$ can act as intermediate propagating baryon with $S=-3$. 
  
\begin{table}[t]
\begin{center}
\resizebox{7.cm}{!} {
\begin{tabular}{c|c|c}
\hline
\hline
$C_{ij}$ & {\bf $K^-\Xi^0$} & {\bf $\bar{K}^0\Xi^-$}  \\  \hline
{\bf $K^-\Xi^0$} & $-1$ & $-1$   \\ 
{\bf $\bar{K}^0\Xi^-$}   &  & $-1$   \\  
\hline
\hline
$D_{ij}$ & {\bf $K^-\Xi^0$} & {\bf $\bar{K}^0\Xi^-$}  \\  \hline
{\bf $K^-\Xi^0$} & $2(2b_0+b_D+b_F)m^{2}_{K}$ & $2(b_D-b_F)m^{2}_{K}$\\ 
{\bf $\bar{K}^0\Xi^-$}   &  & $2(2b_0+b_D+b_F)m^{2}_{K}$   \\   
\hline
\hline
$L_{ij}$ & {\bf $K^-\Xi^0$} & {\bf $\bar{K}^0\Xi^-$}  \\  \hline
{\bf $K^-\Xi^0$} & $d_1+d_2+2d_4$ & $-d_1+d_2+d_3$   \\ 
{\bf $\bar{K}^0\Xi^-$} &  & $d_1+d_2+2d_4$   \\  
\hline
\hline
\end{tabular}
}
\caption{Weinberg--Tomozawa ($C_{ij}$) and NLO ($D_{ij}, L_{ij}$) coefficients of the pseudoscalar meson-ground state baryon interaction kernel in the ($S=-3, Q=-1$) sector. The coefficients are symmetric, $C_{ji} = C_{ij}$, $D_{ji} = D_{ij}$ and $L_{ji} = L_{ij}$. The quantity $m_{K}$ is the average value of the kaon masses.}
\label{coeff_ Qm1}
\end{center}
\end{table}
\begin{table}[t]
\begin{center}
\resizebox{7.cm}{!}{
\begin{tabular}{c|c|c}
\hline
\hline
$C_{K^-\Xi^-,K^-\Xi^-}$ & $D_{K^-\Xi^-,K^-\Xi^-}$ & $L_{K^-\Xi^-,K^-\Xi^-}$  \\  \hline
 $-2$ & \, $4(b_0+b_D)m^{2}_{K}$ \, & \, $2d_2+d_3+2d_4$  \\  
\hline
\hline
$C_{\bar{K}^0\Xi^0,\bar{K}^0\Xi^0}$ & $D_{\bar{K}^0\Xi^0,\bar{K}^0\Xi^0}$ & $L_{\bar{K}^0\Xi^0,\bar{K}^0\Xi^0}$  \\  \hline
 $-2$ & \, $4(b_0+b_D)m^{2}_{K}$ \, & \, $2d_2+d_3+2d_4$  \\  
 \hline
\hline 
\end{tabular}
}
\caption{As Tab.\ \ref{coeff_ Qm1} for the ($S=-3, Q=-2$) and ($S=-3, Q=0$) sectors}. Note that these two sectors have the same coefficients as it should by isospin symmetry, since both are pure $I=1$ channels.
\label{coeff_ Qm2_Q0}
\end{center}
\end{table}

A dynamically generated state shows up as a pole singularity of the scattering amplitude by analytic continuation at a complex energy $\sqrt{s}=z_p=M_R-{\rm i}\Gamma_R/2$, whose real and imaginary parts correspond to its mass ($M_R$) and half width ($\Gamma_R/2$). The complex coupling strengths ($g_i$, $g_j$) of the resonance to the corresponding channels can be evaluated assuming a Breit--Wigner structure for the scattering amplitude in the proximity of the found pole on the real axis, $T_{ij}(\sqrt{s})\sim {g_i g_j} / {(\sqrt{s}-z_p)}$.
\begin{table}[h]
\centering
\resizebox{7.cm}{!} {
\begin{tabular}{c|cc|cc}
\hline
\hline 
BCN Model &    \multicolumn{2}{c|}{$\Omega^*$}    &  \multicolumn{2}{c}{$P_{sss}$} \\
 \hline
$M\;\rm[MeV]$     & \multicolumn{2}{c|}{$2014.06$}   & \multicolumn{2}{c}{$2151.61$}     \\
$\Gamma\;\rm[MeV]$   & \multicolumn{2}{c|}{$296.10$}   & \multicolumn{2}{c}{$399.18$}   \\
\hline
                        &   $ g_i$    &   $|g_i|$          &  $g_i$           & $|g_i|$      \\
$\bar{K} \Xi (I=0)$     &  $1.15+1.78\,i$     &  $2.12$         & $0.00+0.01\,i$    &  $0.01$  \\
$\bar{K}\Xi (I=1)$      &  $-0.00-0.01\,i$    &  $0.01$         & $1.05+1.95\,i$    &  $2.21$  \\
\hline
\hline
 VBC Model &    \multicolumn{2}{c|}{$\Omega^*$}    &  \multicolumn{2}{c}{$P_{sss}$} \\
 \hline
  
$M\;\rm[MeV]$           & \multicolumn{2}{c|}{$1707.24$}      & \multicolumn{2}{c}{$1800.79$}     \\
$\Gamma\;\rm[MeV]$      & \multicolumn{2}{c|}{$0.00$}       & \multicolumn{2}{c}{$0.00$}   \\
\hline
                         &   $ g_i$          &   $|g_i|$      &  $g_i$      & $|g_i|$        \\
$\bar{K} \Xi (I=0)$  &  $4.27+0.00\,i$     &  $4.27$         & $-0.00+0.00\,i$    &  $0.00$  \\
$\bar{K}\Xi (I=1)$          &  $0.01+0.00\,i$     &  $0.01$  & $2.52-0.00\,i$    &  $2.52$  \\
\hline
\hline
\end{tabular}}
  \caption{Pole content with $J^P=\frac{1}{2}^-$  predicted by the two models
  in the ($S=-3, \, Q=-1$) sector. The 
  coupling strengths $g_i$ to the corresponding meson-baryon channels in the isospin basis, and their modulus $|g_i|$, are also shown}.
  \smallskip
  \label{tab:spectroscopy_1}
\end{table}

\begin{table}[h]
\centering
\resizebox{7.cm}{!} {
\begin{tabular}{c|cc|cc}
\hline
\hline 
 $P_{sss}$ &    \multicolumn{2}{c|}{BCN Model}    &  \multicolumn{2}{c}{VBC Model} \\
 \hline
  $M\;\rm[MeV]$           & \multicolumn{2}{c|}{$2152.49$}      & \multicolumn{2}{c}{$1801.47$}     \\
$\Gamma\;\rm[MeV]$      & \multicolumn{2}{c|}{$399.19$}       & \multicolumn{2}{c}{$0.00$}   \\
\hline
&   $ g_i$  &   $|g_i|$  &  $g_i$  & $|g_i|$        \\
$K^- \Xi^-$  &  $1.05+1.95\,i$     &  $2.21$         & $2.54-0.00\,i$    &  $2.54$  \\

\hline
\hline
 $P_{sss}$ &    \multicolumn{2}{c|}{BCN Model}    &  \multicolumn{2}{c}{VBC Model} \\
 \hline
$M\;\rm[MeV]$           & \multicolumn{2}{c|}{$2147.79$}      & \multicolumn{2}{c}{$1800.27$}     \\
$\Gamma\;\rm[MeV]$      & \multicolumn{2}{c|}{$398.76$}       & \multicolumn{2}{c}{$0.00$}   \\
\hline
                         &   $ g_i$          &   $|g_i|$      &  $g_i$      & $|g_i|$        \\
$\bar{K}^0 \Xi^0$  &  $1.05+1.95\,i$     &  $2.22$         & $2.48-0.00\,i$    &  $2.48$  \\
\hline
\hline

\end{tabular}}
\caption{
As Tab.~\ref{tab:spectroscopy_1} for the ($S=-3$, $Q=-2$) and ($S=-3$, $Q=0$) sectors.}
  \smallskip
  \label{tab:spectroscopy_2}
\end{table}

Now, we analyze the pole content of the scattering amplitudes derived from the BCN and the VBC models.
The states found in the ($S=-3,Q=-1$) sector are shown in Tab.\ \ref{tab:spectroscopy_1} whereas those found in the ($S=-3,Q=-2$) and ($S=-3,Q=0$) one are displayed in Tab.\ \ref{tab:spectroscopy_2}.
Focusing first on the ($S=-3,Q=-1$) sector,
where we decide to work in isospin basis to easily differentiate wether the found states are excited $\Omega^*$'s ($I=0$) or pentaquark states $P_{sss}$ ($I=1$), we appreciate that with both models we generate one of each of such structures with spin-parity $J^P=1/2^-$. Within the BCN model, we get a $\Omega^*$ resonance located $342$~MeV above the $\Omega$ ground state that 
could be associated to the $\Omega(2012)$ resonance with unknown spin-parity and three-star status according to the PDG compilation \cite{ParticleDataGroup:2024cfk}. However, this assignment 
can be immediately discarded given the large width obtained theoretically compared to the experimental one, almost with a difference of factor $50$. Moreover, the authors of \cite{Ikeno:2020vqv} provide a plausible explanation for the $\Omega(2012)$ resonance in terms of the molecular picture taking the $\bar{K}\Xi^*$, $\eta \Omega$ and $\bar{K}\Xi$ (d-wave) pairs as CC finding in their approach a very reasonable compatibility with the Belle branching ratio $\Gamma_{\Omega^*\to \pi \bar{K}\Xi}/\Gamma_{\Omega^*\to \bar{K}\Xi}$ \cite{Belle:2019zco}. The $\Omega^*$ generated when employing the VBC model pops up as a bound state, with a binding energy of $\sim 100$~MeV, which precludes its strong decay, thereby leaving processes like $\Omega^* \to \gamma \Omega$ as the most probable decay mechanism. Actually, a $J^P=1/2^-$ virtual state was found in Ref.~\cite{Gamermann:2011mq} with a mass around $1798$~MeV, which is also below the $\bar{K}\Xi$ threshold, and with non-negligible coupling to $\bar{K}\Xi$.

The novelty of the present work comes when inspecting the $I=1$ component in Table~\ref{tab:spectroscopy_1},  
where it can be seen that both models generate a $J^P=1/2^-$ molecular state whose quantum numbers qualify it
as an exotic state. In the case of BCN model, such a pentaquark has a mass of $2152$~MeV and a very large width ($\Gamma=399$~MeV) which makes its potential measurement difficult given the diluted experimental signal in case it existed. In contrast, the VBC model offers a completely different scenario in which a pentaquark is generated approximately $10$~MeV below the $\bar{K}\Xi$ threshold, thus providing a picture that offers greater plausability to the molecular paradigm. These outputs are also reflected in 
the other single channels with charge  $Q=-2$ and $Q=0$, compilated in 
Tab.~\ref{tab:spectroscopy_2}, since they are purely $I=1$.
We note here that the tiny difference observed in the pole location predicted by the two models for the $Q=-2$ and $Q=0$  
charge sectors is due to the explicit use of physical meson and baryon masses. As an illustrative feature of the manifestly exotic nature of the states contained in 
Tab.~\ref{tab:spectroscopy_2}, it is enough to look at their minimal quark-flavor content, being $\ket{\bar{u}dsss}$ for the 
$Q=-2$ charge sector
and $\ket{\bar{d}usss}$ for
the $Q=0$ one.

\begin{figure}[t]
\begin{center}
\includegraphics[width=0.45\textwidth,keepaspectratio]{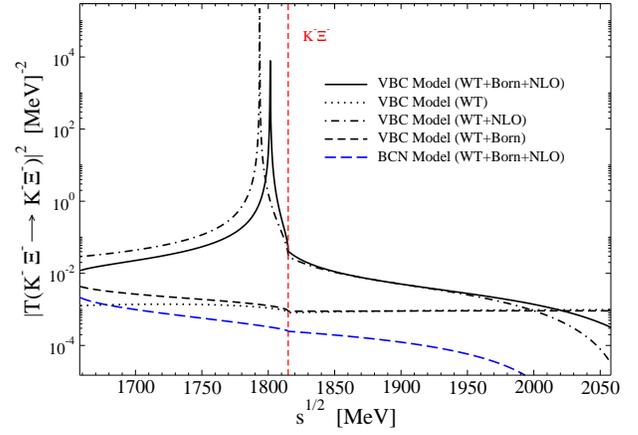}
\caption{(Color on-line) Modulus square of the $K^-\Xi^-$ elastic amplitude for the VBC (solid line) and BCN (long dashed line) models. The dotted, dash-dotted and dashed lines show, in the case of the VBC model, the square value of the amplitude obtained when the interaction kernels is taken as
WT, WT+NLO and WT+Born, respectively. The vertical red dashed line is the $K^-\Xi^-$ threshold.
}
\label{fig:fig2}
\end{center}
\end{figure}
\begin{figure}[t]
\begin{center}
\includegraphics[width=0.45\textwidth,keepaspectratio]{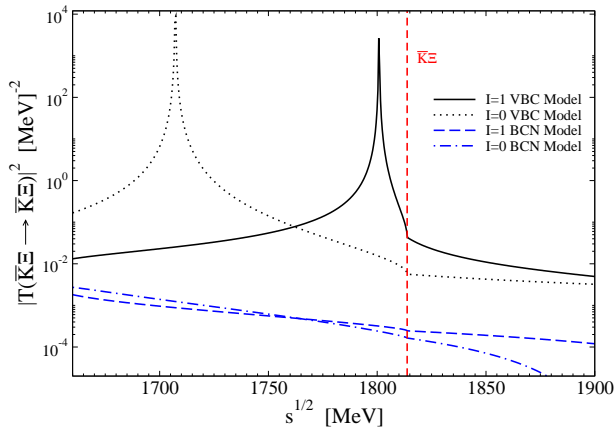}
\caption{ 
(Color-online) Modulus square of the $\bar{K}\Xi$ amplitude in the isospin channels $I=0$ and $I=1$ for the VBC and BCN models. The vertical line indicates the average value of the $\bar{K}\Xi$ threshold.}
\label{fig:fig3}
\end{center}
\end{figure}

Since the scenario drawn by the VBC  
model fits better
with the molecular picture, we focus now more on its results to aid a clearer and more simplified message. In Fig.~\ref{fig:fig2}, the modulus square of the amplitude 
$T(K^-\Xi^-\to K^-\Xi^-)$
is plotted in the energy region where the pentaquark predicted by the
VBC model appears. 
The solid line shows the result obtained when all the contributions to the interaction kernel  are taken into account.
The signal of the $P_{sss}$ state is clearly seen peaking at an energy around $1800$~MeV. The role played by the different building blocks of the interaction kernel obtained with the VBC model
is illustrated by the different lines.
Focusing first on the single WT contribution (dotted line), one can appreciate that its effect is quite shallow. 
Moreover, since the sign of the WT potential 
(see {\it e.g.,} Eq.\ (6) of Ref.\ \cite{Ramos:2016odk}) and that of 
the corresponding coefficient $C_{K^-\Xi^-,K^-\Xi^-}$
(see Table~\ref{coeff_ Qm2_Q0}), are both negative, this contribution is repulsive. The attraction needed to bind the meson-baryon molecule is provided by the NLO contributions, as it can be seen from the dash-dotted line which is obtained when adding them 
to the WT kernel. Being more precise, this effect is due to chiral symmetry breaking pieces accompanying the $b_i$ LECs in Eq.~(\ref{LagrphiB2}).
When comparing the solid line and the dash-dotted black one, it can be inferred that the incorporation of the Born terms (dashed lines) confers an extra repulsion which is not sufficient to dissociate the meson from the baryon in the $P_{sss}$ molecule. This small reduction in the binding energy can be expected from the small difference between the dotted line and dashed one, which translates into a $10$~MeV shift of the lower peak (dashed-dotted line) to the higher energy peak (solid line). We note that in the case of the BCN model (long dashed line) no sign of the $P_{sss}$ state is found in the energy region shown in the figure. The reason is that, in this case, the  amplitude $T(K^-\Xi^-\to K^-\Xi^-)$ is essentially dominated by the repulsive WT term. An additional small repulsion is also due to the contribution of the Born terms, whereas the effect of the attractive NLO ones is negligible, starting to contribute only at energies larger than about 2000 MeV.

The former analysis can be easily generalized for the other two charge sectors. The modulus square of the elastic amplitude $T(\bar{K}^0\Xi^0\to \bar{K}^0\Xi^0)$, not shown for simplicity, is very similar to that of $T(K^-\Xi^-\to K^-\Xi^-)$, as expected 
by isospin symmetry. The small differences observed being just due to the different hadron masses involved in each case. The $Q=-1$ charge sector
has an additional contribution coming from the $I=0$ component of the scattering amplitude, where the interplay among terms entering into the interaction kernel develop a stronger attraction that makes the $\Omega^*$ state arise $100$~MeV below $\bar{K}\Xi$ threshold. This is clearly shown in Fig.~\ref{fig:fig3}, where
$|T(\bar{K}\Xi \to \bar{K}\Xi)|^2$ are depicted in the isospin channels $I=0$ and $I=1$ for both models.
At first glance, it is worth mentioning the noticeable similitude between the $I=1$ modulus square of the amplitude  
and that of $T(K^-\Xi^-\to K^-\Xi^-)$ shown in Fig.~\ref{fig:fig2}. 

Note that if the $P_{sss}$ state exists,
its location and characterizing features 
depend on the parametrization of the UChPT scheme employed. It is therefore difficult to suggest an observable or a process through which this state could be directly observed or inferred, since there are no restrictions on the NLO LECs. However, it is true that measurements from which the $\bar{K}\Xi$ scattering amplitudes can be constrained would provide hints on its existence.
For instance, despite the inherent difficulties in producing the $\Omega_b^-$ baryon, the measurement of the $K^-\Xi^0$ invariant mass distribution associated to the $\Omega_b^-\to J/\psi K^-\Xi^0$ decay, in the LHCb agenda \cite{Misha2023}, can result in a valuable constraint to the NLO driving parameters present in the $I=0$ $K^-\Xi^0$ amplitude \footnote{The $\Omega_b^-\to J/\psi K^-\Xi^0$ decay is an $I=0$ filtering reaction. This can be clearly seen in the formally equivalent $\Lambda_b$ and $\Xi_b$ decays via an internal emission mechanism shown in \cite{Roca:2015tea,Feijoo:2015cca,Feijoo:2024qqg}}. 
Furtheremore,
a direct evidence of the hypothetical $P_{sss}$ signal or, at least, traces of it can come from the experimental measurement of $\bar{K}\Xi$ correlation functions (CFs) which, to the best of our knowledge, are currently being analyzed by the ALICE collaboration.

To illustrate what could be seen in the latter suggested experimental analysis, we calculate the CFs of the $\bar{K}^0\Xi^0$, $K^-\Xi^0$ and $\bar{K}^0\Xi^-$ pairs. The computation of the $K^-\Xi^-$ CF requires the inclusion of the Coulomb interaction in the $T$-matrix. This calculation is more sophisticated and demanding than the one done in the present work since there is not a unique way of treating the Coulomb force, and is left for the near future\footnote{The calculation is currently being undertaken in collaboration with researchers of IFIC (Valencia).}.

\begin{figure}[t!]
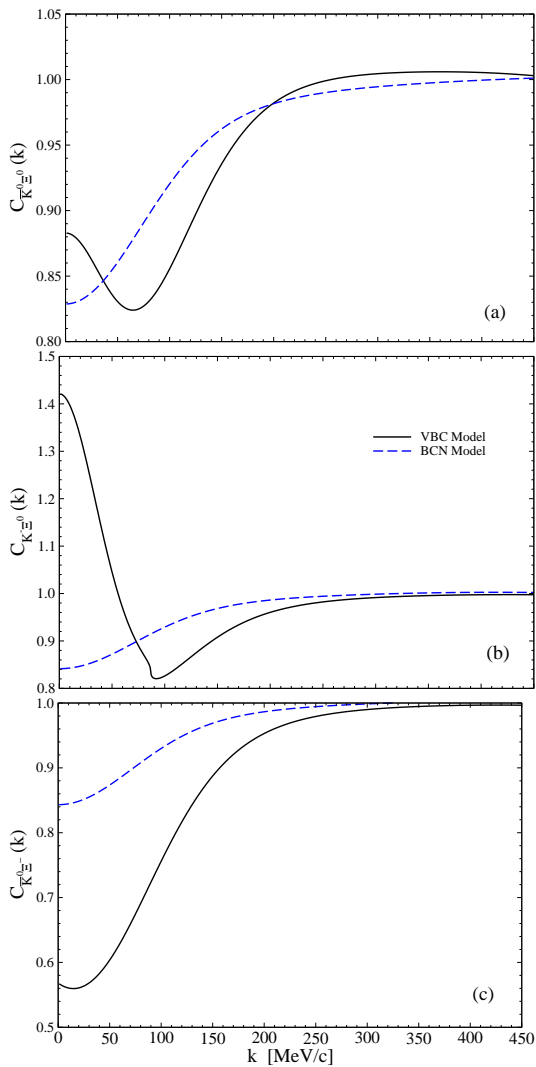

\begin{center}
\includegraphics[width=7.0cm,angle=0,clip]{CF_Q0.eps}
\includegraphics[width=7.0cm,angle=0,clip]{CF_Q1.eps}
\includegraphics[width=7.0cm,angle=0,clip]{CF_Q1b.eps}
\caption{(Color on-line) Correlation functions of the $\bar{K}^0\Xi^0$ (panel a), $K^-\Xi^0$ (panel b) and $\bar{K}^0\Xi^-$ (panel c) pairs predicted by the VBC and BCN models. A emitting source size of $1.1$ fm is assumed.}
\label{fig:CF}
\end{center}
\end{figure}

In the case of a multi-channel system, such as the one we are dealing with in this work, the two-particle CF of a given observed channel $i$ is given by the generalization of the well-known Koonin--Pratt formula 
\cite{Sarti:2023wlg,Koonin:1977fh,Pratt:1986cc,Ohnishi:2016elb,Haidenbauer:2018jvl,Albaladejo:2024lam}
\begin{equation}
C_i(\vec k)=\sum_j\omega_{ji}\int d^3\vec rS_{ji}(\vec r)|\Psi_{ji}(\vec k,\vec r)|^2 \ ,
\label{eq:CF}
\end{equation}
where $\vec k$ is the relative momentum of the particle pair. The contributions of the different channels $j$ are added incoherently and are weighted by the quantities $\omega_{ji}$ determined  from the production yields of the different particles. In the present work we assume 
$\omega_{ji}=1$ for the all the channels involved since their corresponding thresholds differ only by a few MeV.
The function $S_{ji}(\vec r)$ is the so-called source emitting function and, in this work, is taken for all the channels as a spherically symmetric Gaussian distribution normalized to unity. 
The quantity $\Psi_{ji}(\vec k,\vec r)$ 
is the wave function describing the scattering from channel $j$ to channel $i$, and it can be obtained from the scattering amplitude $T_{ji}$ as
\begin{equation}
\Psi_{ji}(\vec k,\vec r)=\delta_{ji}j_0(kr)
+\int d^3\vec q \frac{j_0(qr)T_{ji}(\sqrt{s},\vec k,\vec q)}{\sqrt{s}-E_1^{(j)}(q)-E_2^{(j)}(q)+i\eta} 
\label{eq:WF}
\end{equation}
where $E_1{(j)}(q)$ and $E_2{(q)}$ are the relativistic kinetic energies of particles $1$ and $2$ in channel $j$, $\sqrt{s}=E_1^{(i)}(k)+E_2^{(i)}(k)$ is the total energy of the observed channel $i$ in the CM frame.

The CFs of the pairs $\bar{K}^0\Xi^0$, $K^-\Xi^0$ and $\bar{K}^0\Xi^-$  
predicted by the VBC and BCN models for a emitting source size of $1.1$ fm are shown, respectively, in panels a, b and c of Fig.\ \ref{fig:CF}. In the case of the BCN model the three CFs are smaller than $1$ and increase monotonically to this value, showing this the repulsive character of the interaction kernel of this model. The model VBC, on the other hand, predicts attraction and the existence of the $P_{sss}$ state slightly below the $\bar{K}\Xi$ threshold. This is reflected on the CFs which, for low relative momentum, are larger than $1$, as it is the case of the $K^-\Xi^0$ one, or show the well-known depletion due to the presence of a bound state near threshold if are smaller than $1$, as it is the case of the $\bar{K}^0\Xi^0$ and $\bar{K}^0\Xi^-$ ones.


\textit{Conclusions:} In this letter we have studied the possible existence of a pentaquark $P_{sss}$ state  generated dynamically from the interaction of pseudoscalar mesons with ground state baryons in the strangeness $S=-3$ sector. Collaterally, a molecular $\Omega^*$ have been also found.
We have employed, for the first time in this sector, a unitarized scheme in CC based on the chiral Lagrangian expanded up to NLO. Our results have shown that the inclusion of the NLO terms play a crucial role confering the necessary attractive character to favor the existence of the triply strange pentaquark. The femtoscopic CFs of the $\bar{K}^0\Xi^0$, $K^-\Xi^0$ and $\bar{K}^0\Xi^-$ pairs have been calculated as examples of  possible experimental 
measurements in which a direct evidence of the hypothetical $P_{sss}$ signal could be observed. With the present work, we aim at motivating the experimental analysis of any suggested observable that could confirm or rule out the existence of the proposed pentaquark. Such experimental studies would certainly provide novel insights in this uncharted sector.

\textit{Acknowledgements:}
The authors are very grateful to D. Unkel for the fruitful discussions and their careful reading of the manuscript. A. F. was supported by ORIGINS cluster DFG under Germany’s Excellence Strategy-EXC2094 - 390783311 and the DFG through the Grant SFB 1258 ``Neutrinos and Dark Matter in Astro and Particle Physics”. 


\bibliography{refs.bib}

\begin{thebibliography}{90}
\expandafter\ifx\csname natexlab\endcsname\relax\def\natexlab#1{#1}\fi
\expandafter\ifx\csname bibnamefont\endcsname\relax
  \def\bibnamefont#1{#1}\fi
\expandafter\ifx\csname bibfnamefont\endcsname\relax
  \def\bibfnamefont#1{#1}\fi
\expandafter\ifx\csname citenamefont\endcsname\relax
  \def\citenamefont#1{#1}\fi
\expandafter\ifx\csname url\endcsname\relax
  \def\url#1{\texttt{#1}}\fi
\expandafter\ifx\csname urlprefix\endcsname\relax\def\urlprefix{URL }\fi
\providecommand{\bibinfo}[2]{#2}
\providecommand{\eprint}[2][]{\url{#2}}

\bibitem[{\citenamefont{Aaij et~al.}(2015)}]{LHCb:2015yax}
\bibinfo{author}{\bibfnamefont{R.}~\bibnamefont{Aaij}} \bibnamefont{et~al.}
  (\bibinfo{collaboration}{LHCb}), \bibinfo{journal}{Phys. Rev. Lett.}
  \textbf{\bibinfo{volume}{115}}, \bibinfo{pages}{072001}
  (\bibinfo{year}{2015}), \eprint{1507.03414}.

\bibitem[{\citenamefont{Wu et~al.}(2010)\citenamefont{Wu, Molina, Oset, and
  Zou}}]{Wu:2010jy}
\bibinfo{author}{\bibfnamefont{J.-J.} \bibnamefont{Wu}},
  \bibinfo{author}{\bibfnamefont{R.}~\bibnamefont{Molina}},
  \bibinfo{author}{\bibfnamefont{E.}~\bibnamefont{Oset}}, \bibnamefont{and}
  \bibinfo{author}{\bibfnamefont{B.~S.} \bibnamefont{Zou}},
  \bibinfo{journal}{Phys. Rev. Lett.} \textbf{\bibinfo{volume}{105}},
  \bibinfo{pages}{232001} (\bibinfo{year}{2010}), \eprint{1007.0573}.

\bibitem[{\citenamefont{Wu et~al.}(2011)\citenamefont{Wu, Molina, Oset, and
  Zou}}]{Wu:2010vk}
\bibinfo{author}{\bibfnamefont{J.-J.} \bibnamefont{Wu}},
  \bibinfo{author}{\bibfnamefont{R.}~\bibnamefont{Molina}},
  \bibinfo{author}{\bibfnamefont{E.}~\bibnamefont{Oset}}, \bibnamefont{and}
  \bibinfo{author}{\bibfnamefont{B.~S.} \bibnamefont{Zou}},
  \bibinfo{journal}{Phys. Rev. C} \textbf{\bibinfo{volume}{84}},
  \bibinfo{pages}{015202} (\bibinfo{year}{2011}), \eprint{1011.2399}.

\bibitem[{\citenamefont{Yang et~al.}(2012)\citenamefont{Yang, Sun, He, Liu, and
  Zhu}}]{Yang:2011wz}
\bibinfo{author}{\bibfnamefont{Z.-C.} \bibnamefont{Yang}},
  \bibinfo{author}{\bibfnamefont{Z.-F.} \bibnamefont{Sun}},
  \bibinfo{author}{\bibfnamefont{J.}~\bibnamefont{He}},
  \bibinfo{author}{\bibfnamefont{X.}~\bibnamefont{Liu}}, \bibnamefont{and}
  \bibinfo{author}{\bibfnamefont{S.-L.} \bibnamefont{Zhu}},
  \bibinfo{journal}{Chin. Phys. C} \textbf{\bibinfo{volume}{36}},
  \bibinfo{pages}{6} (\bibinfo{year}{2012}), \eprint{1105.2901}.

\bibitem[{\citenamefont{Xiao et~al.}(2013)\citenamefont{Xiao, Nieves, and
  Oset}}]{Xiao:2013yca}
\bibinfo{author}{\bibfnamefont{C.~W.} \bibnamefont{Xiao}},
  \bibinfo{author}{\bibfnamefont{J.}~\bibnamefont{Nieves}}, \bibnamefont{and}
  \bibinfo{author}{\bibfnamefont{E.}~\bibnamefont{Oset}},
  \bibinfo{journal}{Phys. Rev. D} \textbf{\bibinfo{volume}{88}},
  \bibinfo{pages}{056012} (\bibinfo{year}{2013}), \eprint{1304.5368}.

\bibitem[{\citenamefont{Karliner and Rosner}(2015)}]{Karliner:2015ina}
\bibinfo{author}{\bibfnamefont{M.}~\bibnamefont{Karliner}} \bibnamefont{and}
  \bibinfo{author}{\bibfnamefont{J.~L.} \bibnamefont{Rosner}},
  \bibinfo{journal}{Phys. Rev. Lett.} \textbf{\bibinfo{volume}{115}},
  \bibinfo{pages}{122001} (\bibinfo{year}{2015}), \eprint{1506.06386}.

\bibitem[{\citenamefont{Wang et~al.}(2011)\citenamefont{Wang, Huang, Zhang, and
  Zou}}]{Wang:2011rga}
\bibinfo{author}{\bibfnamefont{W.~L.} \bibnamefont{Wang}},
  \bibinfo{author}{\bibfnamefont{F.}~\bibnamefont{Huang}},
  \bibinfo{author}{\bibfnamefont{Z.~Y.} \bibnamefont{Zhang}}, \bibnamefont{and}
  \bibinfo{author}{\bibfnamefont{B.~S.} \bibnamefont{Zou}},
  \bibinfo{journal}{Phys. Rev. C} \textbf{\bibinfo{volume}{84}},
  \bibinfo{pages}{015203} (\bibinfo{year}{2011}), \eprint{1101.0453}.

\bibitem[{\citenamefont{Yuan et~al.}(2012)\citenamefont{Yuan, Wei, He, Xu, and
  Zou}}]{Yuan:2012wz}
\bibinfo{author}{\bibfnamefont{S.~G.} \bibnamefont{Yuan}},
  \bibinfo{author}{\bibfnamefont{K.~W.} \bibnamefont{Wei}},
  \bibinfo{author}{\bibfnamefont{J.}~\bibnamefont{He}},
  \bibinfo{author}{\bibfnamefont{H.~S.} \bibnamefont{Xu}}, \bibnamefont{and}
  \bibinfo{author}{\bibfnamefont{B.~S.} \bibnamefont{Zou}},
  \bibinfo{journal}{Eur. Phys. J. A} \textbf{\bibinfo{volume}{48}},
  \bibinfo{pages}{61} (\bibinfo{year}{2012}), \eprint{1201.0807}.

\bibitem[{\citenamefont{Aaij et~al.}(2022)}]{LHCb:2021chn}
\bibinfo{author}{\bibfnamefont{R.}~\bibnamefont{Aaij}} \bibnamefont{et~al.}
  (\bibinfo{collaboration}{LHCb}), \bibinfo{journal}{Phys. Rev. Lett.}
  \textbf{\bibinfo{volume}{128}}, \bibinfo{pages}{062001}
  (\bibinfo{year}{2022}), \eprint{2108.04720}.

\bibitem[{\citenamefont{Chen et~al.}(2015{\natexlab{a}})\citenamefont{Chen,
  Liu, Li, and Zhu}}]{Chen:2015loa}
\bibinfo{author}{\bibfnamefont{R.}~\bibnamefont{Chen}},
  \bibinfo{author}{\bibfnamefont{X.}~\bibnamefont{Liu}},
  \bibinfo{author}{\bibfnamefont{X.-Q.} \bibnamefont{Li}}, \bibnamefont{and}
  \bibinfo{author}{\bibfnamefont{S.-L.} \bibnamefont{Zhu}},
  \bibinfo{journal}{Phys. Rev. Lett.} \textbf{\bibinfo{volume}{115}},
  \bibinfo{pages}{132002} (\bibinfo{year}{2015}{\natexlab{a}}),
  \eprint{1507.03704}.

\bibitem[{\citenamefont{He}(2016)}]{He:2015cea}
\bibinfo{author}{\bibfnamefont{J.}~\bibnamefont{He}}, \bibinfo{journal}{Phys.
  Lett. B} \textbf{\bibinfo{volume}{753}}, \bibinfo{pages}{547}
  (\bibinfo{year}{2016}), \eprint{1507.05200}.

\bibitem[{\citenamefont{Liu et~al.}(2019)\citenamefont{Liu, Pan, Peng,
  S\'anchez~S\'anchez, Geng, Hosaka, and Pavon~Valderrama}}]{Liu:2019tjn}
\bibinfo{author}{\bibfnamefont{M.-Z.} \bibnamefont{Liu}},
  \bibinfo{author}{\bibfnamefont{Y.-W.} \bibnamefont{Pan}},
  \bibinfo{author}{\bibfnamefont{F.-Z.} \bibnamefont{Peng}},
  \bibinfo{author}{\bibfnamefont{M.}~\bibnamefont{S\'anchez~S\'anchez}},
  \bibinfo{author}{\bibfnamefont{L.-S.} \bibnamefont{Geng}},
  \bibinfo{author}{\bibfnamefont{A.}~\bibnamefont{Hosaka}}, \bibnamefont{and}
  \bibinfo{author}{\bibfnamefont{M.}~\bibnamefont{Pavon~Valderrama}},
  \bibinfo{journal}{Phys. Rev. Lett.} \textbf{\bibinfo{volume}{122}},
  \bibinfo{pages}{242001} (\bibinfo{year}{2019}), \eprint{1903.11560}.

\bibitem[{\citenamefont{Du et~al.}(2021)\citenamefont{Du, Baru, Guo, Hanhart,
  Mei\ss{}ner, Oller, and Wang}}]{Du:2021fmf}
\bibinfo{author}{\bibfnamefont{M.-L.} \bibnamefont{Du}},
  \bibinfo{author}{\bibfnamefont{V.}~\bibnamefont{Baru}},
  \bibinfo{author}{\bibfnamefont{F.-K.} \bibnamefont{Guo}},
  \bibinfo{author}{\bibfnamefont{C.}~\bibnamefont{Hanhart}},
  \bibinfo{author}{\bibfnamefont{U.-G.} \bibnamefont{Mei\ss{}ner}},
  \bibinfo{author}{\bibfnamefont{J.~A.} \bibnamefont{Oller}}, \bibnamefont{and}
  \bibinfo{author}{\bibfnamefont{Q.}~\bibnamefont{Wang}},
  \bibinfo{journal}{JHEP} \textbf{\bibinfo{volume}{08}}, \bibinfo{pages}{157}
  (\bibinfo{year}{2021}), \eprint{2102.07159}.

\bibitem[{\citenamefont{Chen et~al.}(2015{\natexlab{b}})\citenamefont{Chen,
  Chen, Liu, Steele, and Zhu}}]{Chen:2015moa}
\bibinfo{author}{\bibfnamefont{H.-X.} \bibnamefont{Chen}},
  \bibinfo{author}{\bibfnamefont{W.}~\bibnamefont{Chen}},
  \bibinfo{author}{\bibfnamefont{X.}~\bibnamefont{Liu}},
  \bibinfo{author}{\bibfnamefont{T.~G.} \bibnamefont{Steele}},
  \bibnamefont{and} \bibinfo{author}{\bibfnamefont{S.-L.} \bibnamefont{Zhu}},
  \bibinfo{journal}{Phys. Rev. Lett.} \textbf{\bibinfo{volume}{115}},
  \bibinfo{pages}{172001} (\bibinfo{year}{2015}{\natexlab{b}}),
  \eprint{1507.03717}.

\bibitem[{\citenamefont{Wang}(2016{\natexlab{a}})}]{Wang:2015epa}
\bibinfo{author}{\bibfnamefont{Z.-G.} \bibnamefont{Wang}},
  \bibinfo{journal}{Eur. Phys. J. C} \textbf{\bibinfo{volume}{76}},
  \bibinfo{pages}{70} (\bibinfo{year}{2016}{\natexlab{a}}),
  \eprint{1508.01468}.

\bibitem[{\citenamefont{Wang}(2020)}]{Wang:2019got}
\bibinfo{author}{\bibfnamefont{Z.-G.} \bibnamefont{Wang}},
  \bibinfo{journal}{Int. J. Mod. Phys. A} \textbf{\bibinfo{volume}{35}},
  \bibinfo{pages}{2050003} (\bibinfo{year}{2020}), \eprint{1905.02892}.

\bibitem[{\citenamefont{Ortega et~al.}(2017)\citenamefont{Ortega, Entem, and
  Fern\'andez}}]{Ortega:2016syt}
\bibinfo{author}{\bibfnamefont{P.~G.} \bibnamefont{Ortega}},
  \bibinfo{author}{\bibfnamefont{D.~R.} \bibnamefont{Entem}}, \bibnamefont{and}
  \bibinfo{author}{\bibfnamefont{F.}~\bibnamefont{Fern\'andez}},
  \bibinfo{journal}{Phys. Lett. B} \textbf{\bibinfo{volume}{764}},
  \bibinfo{pages}{207} (\bibinfo{year}{2017}), \eprint{1606.06148}.

\bibitem[{\citenamefont{Park et~al.}(2017)\citenamefont{Park, Park, Cho, and
  Lee}}]{Park:2017jbn}
\bibinfo{author}{\bibfnamefont{W.}~\bibnamefont{Park}},
  \bibinfo{author}{\bibfnamefont{A.}~\bibnamefont{Park}},
  \bibinfo{author}{\bibfnamefont{S.}~\bibnamefont{Cho}}, \bibnamefont{and}
  \bibinfo{author}{\bibfnamefont{S.~H.} \bibnamefont{Lee}},
  \bibinfo{journal}{Phys. Rev. D} \textbf{\bibinfo{volume}{95}},
  \bibinfo{pages}{054027} (\bibinfo{year}{2017}), \eprint{1702.00381}.

\bibitem[{\citenamefont{Weng et~al.}(2019)\citenamefont{Weng, Chen, Deng, and
  Zhu}}]{Weng:2019ynv}
\bibinfo{author}{\bibfnamefont{X.-Z.} \bibnamefont{Weng}},
  \bibinfo{author}{\bibfnamefont{X.-L.} \bibnamefont{Chen}},
  \bibinfo{author}{\bibfnamefont{W.-Z.} \bibnamefont{Deng}}, \bibnamefont{and}
  \bibinfo{author}{\bibfnamefont{S.-L.} \bibnamefont{Zhu}},
  \bibinfo{journal}{Phys. Rev. D} \textbf{\bibinfo{volume}{100}},
  \bibinfo{pages}{016014} (\bibinfo{year}{2019}), \eprint{1904.09891}.

\bibitem[{\citenamefont{Zhu et~al.}(2019)\citenamefont{Zhu, Liu, Huang, and
  Qiao}}]{Zhu:2019iwm}
\bibinfo{author}{\bibfnamefont{R.}~\bibnamefont{Zhu}},
  \bibinfo{author}{\bibfnamefont{X.}~\bibnamefont{Liu}},
  \bibinfo{author}{\bibfnamefont{H.}~\bibnamefont{Huang}}, \bibnamefont{and}
  \bibinfo{author}{\bibfnamefont{C.-F.} \bibnamefont{Qiao}},
  \bibinfo{journal}{Phys. Lett. B} \textbf{\bibinfo{volume}{797}},
  \bibinfo{pages}{134869} (\bibinfo{year}{2019}), \eprint{1904.10285}.

\bibitem[{\citenamefont{Deng}(2022)}]{Deng:2022vkv}
\bibinfo{author}{\bibfnamefont{C.-R.} \bibnamefont{Deng}},
  \bibinfo{journal}{Phys. Rev. D} \textbf{\bibinfo{volume}{105}},
  \bibinfo{pages}{116021} (\bibinfo{year}{2022}), \eprint{2202.13570}.

\bibitem[{\citenamefont{Guo et~al.}(2015)\citenamefont{Guo, Mei\ss{}ner, Wang,
  and Yang}}]{Guo:2015umn}
\bibinfo{author}{\bibfnamefont{F.-K.} \bibnamefont{Guo}},
  \bibinfo{author}{\bibfnamefont{U.-G.} \bibnamefont{Mei\ss{}ner}},
  \bibinfo{author}{\bibfnamefont{W.}~\bibnamefont{Wang}}, \bibnamefont{and}
  \bibinfo{author}{\bibfnamefont{Z.}~\bibnamefont{Yang}},
  \bibinfo{journal}{Phys. Rev. D} \textbf{\bibinfo{volume}{92}},
  \bibinfo{pages}{071502} (\bibinfo{year}{2015}), \eprint{1507.04950}.

\bibitem[{\citenamefont{Liu et~al.}(2016)\citenamefont{Liu, Wang, and
  Zhao}}]{Liu:2015fea}
\bibinfo{author}{\bibfnamefont{X.-H.} \bibnamefont{Liu}},
  \bibinfo{author}{\bibfnamefont{Q.}~\bibnamefont{Wang}}, \bibnamefont{and}
  \bibinfo{author}{\bibfnamefont{Q.}~\bibnamefont{Zhao}},
  \bibinfo{journal}{Phys. Lett. B} \textbf{\bibinfo{volume}{757}},
  \bibinfo{pages}{231} (\bibinfo{year}{2016}), \eprint{1507.05359}.

\bibitem[{\citenamefont{Mikhasenko}(2015)}]{Mikhasenko:2015vca}
\bibinfo{author}{\bibfnamefont{M.}~\bibnamefont{Mikhasenko}}
  (\bibinfo{year}{2015}), \eprint{1507.06552}.

\bibitem[{\citenamefont{Nakamura et~al.}(2021)\citenamefont{Nakamura, Hosaka,
  and Yamaguchi}}]{Nakamura:2021dix}
\bibinfo{author}{\bibfnamefont{S.~X.} \bibnamefont{Nakamura}},
  \bibinfo{author}{\bibfnamefont{A.}~\bibnamefont{Hosaka}}, \bibnamefont{and}
  \bibinfo{author}{\bibfnamefont{Y.}~\bibnamefont{Yamaguchi}},
  \bibinfo{journal}{Phys. Rev. D} \textbf{\bibinfo{volume}{104}},
  \bibinfo{pages}{L091503} (\bibinfo{year}{2021}), \eprint{2109.15235}.

\bibitem[{\citenamefont{Hofmann and Lutz}(2005)}]{Hofmann:2005sw}
\bibinfo{author}{\bibfnamefont{J.}~\bibnamefont{Hofmann}} \bibnamefont{and}
  \bibinfo{author}{\bibfnamefont{M.~F.~M.} \bibnamefont{Lutz}},
  \bibinfo{journal}{Nucl. Phys. A} \textbf{\bibinfo{volume}{763}},
  \bibinfo{pages}{90} (\bibinfo{year}{2005}), \eprint{hep-ph/0507071}.

\bibitem[{\citenamefont{Wang et~al.}(2020)\citenamefont{Wang, Meng, and
  Zhu}}]{Wang:2019nvm}
\bibinfo{author}{\bibfnamefont{B.}~\bibnamefont{Wang}},
  \bibinfo{author}{\bibfnamefont{L.}~\bibnamefont{Meng}}, \bibnamefont{and}
  \bibinfo{author}{\bibfnamefont{S.-L.} \bibnamefont{Zhu}},
  \bibinfo{journal}{Phys. Rev. D} \textbf{\bibinfo{volume}{101}},
  \bibinfo{pages}{034018} (\bibinfo{year}{2020}), \eprint{1912.12592}.

\bibitem[{\citenamefont{Anisovich et~al.}(2015)\citenamefont{Anisovich,
  Matveev, Nyiri, Sarantsev, and Semenova}}]{Anisovich:2015zqa}
\bibinfo{author}{\bibfnamefont{V.~V.} \bibnamefont{Anisovich}},
  \bibinfo{author}{\bibfnamefont{M.~A.} \bibnamefont{Matveev}},
  \bibinfo{author}{\bibfnamefont{J.}~\bibnamefont{Nyiri}},
  \bibinfo{author}{\bibfnamefont{A.~V.} \bibnamefont{Sarantsev}},
  \bibnamefont{and} \bibinfo{author}{\bibfnamefont{A.~N.}
  \bibnamefont{Semenova}}, \bibinfo{journal}{Int. J. Mod. Phys. A}
  \textbf{\bibinfo{volume}{30}}, \bibinfo{pages}{1550190}
  (\bibinfo{year}{2015}), \eprint{1509.04898}.

\bibitem[{\citenamefont{Wang}(2016{\natexlab{b}})}]{Wang:2015wsa}
\bibinfo{author}{\bibfnamefont{Z.-G.} \bibnamefont{Wang}},
  \bibinfo{journal}{Eur. Phys. J. C} \textbf{\bibinfo{volume}{76}},
  \bibinfo{pages}{142} (\bibinfo{year}{2016}{\natexlab{b}}),
  \eprint{1509.06436}.

\bibitem[{\citenamefont{Feijoo et~al.}(2016)\citenamefont{Feijoo, Magas, Ramos,
  and Oset}}]{Feijoo:2015kts}
\bibinfo{author}{\bibfnamefont{A.}~\bibnamefont{Feijoo}},
  \bibinfo{author}{\bibfnamefont{V.~K.} \bibnamefont{Magas}},
  \bibinfo{author}{\bibfnamefont{A.}~\bibnamefont{Ramos}}, \bibnamefont{and}
  \bibinfo{author}{\bibfnamefont{E.}~\bibnamefont{Oset}},
  \bibinfo{journal}{Eur. Phys. J. C} \textbf{\bibinfo{volume}{76}},
  \bibinfo{pages}{446} (\bibinfo{year}{2016}), \eprint{1512.08152}.

\bibitem[{\citenamefont{Lu et~al.}(2016)\citenamefont{Lu, Wang, Xie, Geng, and
  Oset}}]{Lu:2016roh}
\bibinfo{author}{\bibfnamefont{J.-X.} \bibnamefont{Lu}},
  \bibinfo{author}{\bibfnamefont{E.}~\bibnamefont{Wang}},
  \bibinfo{author}{\bibfnamefont{J.-J.} \bibnamefont{Xie}},
  \bibinfo{author}{\bibfnamefont{L.-S.} \bibnamefont{Geng}}, \bibnamefont{and}
  \bibinfo{author}{\bibfnamefont{E.}~\bibnamefont{Oset}},
  \bibinfo{journal}{Phys. Rev. D} \textbf{\bibinfo{volume}{93}},
  \bibinfo{pages}{094009} (\bibinfo{year}{2016}), \eprint{1601.00075}.

\bibitem[{\citenamefont{Chen et~al.}(2016)\citenamefont{Chen, Geng, Liang,
  Oset, Wang, and Xie}}]{Chen:2015sxa}
\bibinfo{author}{\bibfnamefont{H.-X.} \bibnamefont{Chen}},
  \bibinfo{author}{\bibfnamefont{L.-S.} \bibnamefont{Geng}},
  \bibinfo{author}{\bibfnamefont{W.-H.} \bibnamefont{Liang}},
  \bibinfo{author}{\bibfnamefont{E.}~\bibnamefont{Oset}},
  \bibinfo{author}{\bibfnamefont{E.}~\bibnamefont{Wang}}, \bibnamefont{and}
  \bibinfo{author}{\bibfnamefont{J.-J.} \bibnamefont{Xie}},
  \bibinfo{journal}{Phys. Rev. C} \textbf{\bibinfo{volume}{93}},
  \bibinfo{pages}{065203} (\bibinfo{year}{2016}), \eprint{1510.01803}.

\bibitem[{\citenamefont{Shen et~al.}(2020)\citenamefont{Shen, Jing, Guo, and
  Wu}}]{Shen:2020gpw}
\bibinfo{author}{\bibfnamefont{C.-W.} \bibnamefont{Shen}},
  \bibinfo{author}{\bibfnamefont{H.-J.} \bibnamefont{Jing}},
  \bibinfo{author}{\bibfnamefont{F.-K.} \bibnamefont{Guo}}, \bibnamefont{and}
  \bibinfo{author}{\bibfnamefont{J.-J.} \bibnamefont{Wu}},
  \bibinfo{journal}{Symmetry} \textbf{\bibinfo{volume}{12}},
  \bibinfo{pages}{1611} (\bibinfo{year}{2020}), \eprint{2008.09082}.

\bibitem[{\citenamefont{Aaij et~al.}(2021)}]{LHCb:2020jpq}
\bibinfo{author}{\bibfnamefont{R.}~\bibnamefont{Aaij}} \bibnamefont{et~al.}
  (\bibinfo{collaboration}{LHCb}), \bibinfo{journal}{Sci. Bull.}
  \textbf{\bibinfo{volume}{66}}, \bibinfo{pages}{1278} (\bibinfo{year}{2021}),
  \eprint{2012.10380}.

\bibitem[{\citenamefont{Aaij et~al.}(2023)}]{LHCb:2022ogu}
\bibinfo{author}{\bibfnamefont{R.}~\bibnamefont{Aaij}} \bibnamefont{et~al.}
  (\bibinfo{collaboration}{LHCb}), \bibinfo{journal}{Phys. Rev. Lett.}
  \textbf{\bibinfo{volume}{131}}, \bibinfo{pages}{031901}
  (\bibinfo{year}{2023}), \eprint{2210.10346}.

\bibitem[{\citenamefont{Chen et~al.}(2021)\citenamefont{Chen, Chen, Liu, and
  Liu}}]{Chen:2020uif}
\bibinfo{author}{\bibfnamefont{H.-X.} \bibnamefont{Chen}},
  \bibinfo{author}{\bibfnamefont{W.}~\bibnamefont{Chen}},
  \bibinfo{author}{\bibfnamefont{X.}~\bibnamefont{Liu}}, \bibnamefont{and}
  \bibinfo{author}{\bibfnamefont{X.-H.} \bibnamefont{Liu}},
  \bibinfo{journal}{Eur. Phys. J. C} \textbf{\bibinfo{volume}{81}},
  \bibinfo{pages}{409} (\bibinfo{year}{2021}), \eprint{2011.01079}.

\bibitem[{\citenamefont{Chen}(2022)}]{Chen:2020opr}
\bibinfo{author}{\bibfnamefont{H.-X.} \bibnamefont{Chen}},
  \bibinfo{journal}{Chin. Phys. C} \textbf{\bibinfo{volume}{46}},
  \bibinfo{pages}{093105} (\bibinfo{year}{2022}), \eprint{2011.07187}.

\bibitem[{\citenamefont{Liu et~al.}(2021)\citenamefont{Liu, Pan, and
  Geng}}]{Liu:2020hcv}
\bibinfo{author}{\bibfnamefont{M.-Z.} \bibnamefont{Liu}},
  \bibinfo{author}{\bibfnamefont{Y.-W.} \bibnamefont{Pan}}, \bibnamefont{and}
  \bibinfo{author}{\bibfnamefont{L.-S.} \bibnamefont{Geng}},
  \bibinfo{journal}{Phys. Rev. D} \textbf{\bibinfo{volume}{103}},
  \bibinfo{pages}{034003} (\bibinfo{year}{2021}), \eprint{2011.07935}.

\bibitem[{\citenamefont{Feijoo et~al.}(2023{\natexlab{a}})\citenamefont{Feijoo,
  Wang, Xiao, Wu, Oset, Nieves, and Zou}}]{Feijoo:2022rxf}
\bibinfo{author}{\bibfnamefont{A.}~\bibnamefont{Feijoo}},
  \bibinfo{author}{\bibfnamefont{W.-F.} \bibnamefont{Wang}},
  \bibinfo{author}{\bibfnamefont{C.-W.} \bibnamefont{Xiao}},
  \bibinfo{author}{\bibfnamefont{J.-J.} \bibnamefont{Wu}},
  \bibinfo{author}{\bibfnamefont{E.}~\bibnamefont{Oset}},
  \bibinfo{author}{\bibfnamefont{J.}~\bibnamefont{Nieves}}, \bibnamefont{and}
  \bibinfo{author}{\bibfnamefont{B.-S.} \bibnamefont{Zou}},
  \bibinfo{journal}{Phys. Lett. B} \textbf{\bibinfo{volume}{839}},
  \bibinfo{pages}{137760} (\bibinfo{year}{2023}{\natexlab{a}}),
  \eprint{2212.12223}.

\bibitem[{\citenamefont{Karliner and Rosner}(2022)}]{Karliner:2022erb}
\bibinfo{author}{\bibfnamefont{M.}~\bibnamefont{Karliner}} \bibnamefont{and}
  \bibinfo{author}{\bibfnamefont{J.~L.} \bibnamefont{Rosner}},
  \bibinfo{journal}{Phys. Rev. D} \textbf{\bibinfo{volume}{106}},
  \bibinfo{pages}{036024} (\bibinfo{year}{2022}), \eprint{2207.07581}.

\bibitem[{\citenamefont{Wang and Liu}(2022)}]{Wang:2022mxy}
\bibinfo{author}{\bibfnamefont{F.-L.} \bibnamefont{Wang}} \bibnamefont{and}
  \bibinfo{author}{\bibfnamefont{X.}~\bibnamefont{Liu}},
  \bibinfo{journal}{Phys. Lett. B} \textbf{\bibinfo{volume}{835}},
  \bibinfo{pages}{137583} (\bibinfo{year}{2022}), \eprint{2207.10493}.

\bibitem[{\citenamefont{Yan et~al.}(2023)\citenamefont{Yan, Peng,
  S\'anchez~S\'anchez, and Pavon~Valderrama}}]{Yan:2022wuz}
\bibinfo{author}{\bibfnamefont{M.-J.} \bibnamefont{Yan}},
  \bibinfo{author}{\bibfnamefont{F.-Z.} \bibnamefont{Peng}},
  \bibinfo{author}{\bibfnamefont{M.}~\bibnamefont{S\'anchez~S\'anchez}},
  \bibnamefont{and}
  \bibinfo{author}{\bibfnamefont{M.}~\bibnamefont{Pavon~Valderrama}},
  \bibinfo{journal}{Phys. Rev. D} \textbf{\bibinfo{volume}{107}},
  \bibinfo{pages}{074025} (\bibinfo{year}{2023}), \eprint{2207.11144}.

\bibitem[{\citenamefont{\"Ozdem}(2023)}]{Ozdem:2022kei}
\bibinfo{author}{\bibfnamefont{U.}~\bibnamefont{\"Ozdem}},
  \bibinfo{journal}{Phys. Lett. B} \textbf{\bibinfo{volume}{836}},
  \bibinfo{pages}{137635} (\bibinfo{year}{2023}), \eprint{2208.07684}.

\bibitem[{\citenamefont{Ortega et~al.}(2023)\citenamefont{Ortega, Entem, and
  Fernandez}}]{Ortega:2022uyu}
\bibinfo{author}{\bibfnamefont{P.~G.} \bibnamefont{Ortega}},
  \bibinfo{author}{\bibfnamefont{D.~R.} \bibnamefont{Entem}}, \bibnamefont{and}
  \bibinfo{author}{\bibfnamefont{F.}~\bibnamefont{Fernandez}},
  \bibinfo{journal}{Phys. Lett. B} \textbf{\bibinfo{volume}{838}},
  \bibinfo{pages}{137747} (\bibinfo{year}{2023}), \eprint{2210.04465}.

\bibitem[{\citenamefont{Wang et~al.}(2022)\citenamefont{Wang, Zhou, Liu, and
  Liu}}]{Wang:2022tib}
\bibinfo{author}{\bibfnamefont{F.-L.} \bibnamefont{Wang}},
  \bibinfo{author}{\bibfnamefont{H.-Y.} \bibnamefont{Zhou}},
  \bibinfo{author}{\bibfnamefont{Z.-W.} \bibnamefont{Liu}}, \bibnamefont{and}
  \bibinfo{author}{\bibfnamefont{X.}~\bibnamefont{Liu}},
  \bibinfo{journal}{Phys. Rev. D} \textbf{\bibinfo{volume}{106}},
  \bibinfo{pages}{054020} (\bibinfo{year}{2022}), \eprint{2208.10756}.

\bibitem[{\citenamefont{Peng et~al.}(2021)\citenamefont{Peng, Yan,
  S\'anchez~S\'anchez, and Valderrama}}]{Peng:2020hql}
\bibinfo{author}{\bibfnamefont{F.-Z.} \bibnamefont{Peng}},
  \bibinfo{author}{\bibfnamefont{M.-J.} \bibnamefont{Yan}},
  \bibinfo{author}{\bibfnamefont{M.}~\bibnamefont{S\'anchez~S\'anchez}},
  \bibnamefont{and} \bibinfo{author}{\bibfnamefont{M.~P.}
  \bibnamefont{Valderrama}}, \bibinfo{journal}{Eur. Phys. J. C}
  \textbf{\bibinfo{volume}{81}}, \bibinfo{pages}{666} (\bibinfo{year}{2021}),
  \eprint{2011.01915}.

\bibitem[{\citenamefont{Xiao et~al.}(2021)\citenamefont{Xiao, Wu, and
  Zou}}]{Xiao:2021rgp}
\bibinfo{author}{\bibfnamefont{C.~W.} \bibnamefont{Xiao}},
  \bibinfo{author}{\bibfnamefont{J.~J.} \bibnamefont{Wu}}, \bibnamefont{and}
  \bibinfo{author}{\bibfnamefont{B.~S.} \bibnamefont{Zou}},
  \bibinfo{journal}{Phys. Rev. D} \textbf{\bibinfo{volume}{103}},
  \bibinfo{pages}{054016} (\bibinfo{year}{2021}), \eprint{2102.02607}.

\bibitem[{\citenamefont{Zhu et~al.}(2021)\citenamefont{Zhu, Song, and
  He}}]{Zhu:2021lhd}
\bibinfo{author}{\bibfnamefont{J.-T.} \bibnamefont{Zhu}},
  \bibinfo{author}{\bibfnamefont{L.-Q.} \bibnamefont{Song}}, \bibnamefont{and}
  \bibinfo{author}{\bibfnamefont{J.}~\bibnamefont{He}}, \bibinfo{journal}{Phys.
  Rev. D} \textbf{\bibinfo{volume}{103}}, \bibinfo{pages}{074007}
  (\bibinfo{year}{2021}), \eprint{2101.12441}.

\bibitem[{\citenamefont{Burns and Swanson}(2023)}]{Burns:2022uha}
\bibinfo{author}{\bibfnamefont{T.~J.} \bibnamefont{Burns}} \bibnamefont{and}
  \bibinfo{author}{\bibfnamefont{E.~S.} \bibnamefont{Swanson}},
  \bibinfo{journal}{Phys. Lett. B} \textbf{\bibinfo{volume}{838}},
  \bibinfo{pages}{137715} (\bibinfo{year}{2023}), \eprint{2208.05106}.

\bibitem[{\citenamefont{Nakamura and Wu}(2023)}]{Nakamura:2022gtu}
\bibinfo{author}{\bibfnamefont{S.~X.} \bibnamefont{Nakamura}} \bibnamefont{and}
  \bibinfo{author}{\bibfnamefont{J.~J.} \bibnamefont{Wu}},
  \bibinfo{journal}{Phys. Rev. D} \textbf{\bibinfo{volume}{108}},
  \bibinfo{pages}{L011501} (\bibinfo{year}{2023}), \eprint{2208.11995}.

\bibitem[{\citenamefont{Meng et~al.}(2023)\citenamefont{Meng, Wang, and
  Zhu}}]{Meng:2022wgl}
\bibinfo{author}{\bibfnamefont{L.}~\bibnamefont{Meng}},
  \bibinfo{author}{\bibfnamefont{B.}~\bibnamefont{Wang}}, \bibnamefont{and}
  \bibinfo{author}{\bibfnamefont{S.-L.} \bibnamefont{Zhu}},
  \bibinfo{journal}{Phys. Rev. D} \textbf{\bibinfo{volume}{107}},
  \bibinfo{pages}{014005} (\bibinfo{year}{2023}), \eprint{2208.03883}.

\bibitem[{\citenamefont{Wu and Chen}(2024)}]{Wu:2024lud}
\bibinfo{author}{\bibfnamefont{Q.}~\bibnamefont{Wu}} \bibnamefont{and}
  \bibinfo{author}{\bibfnamefont{D.-Y.} \bibnamefont{Chen}},
  \bibinfo{journal}{Phys. Rev. D} \textbf{\bibinfo{volume}{109}},
  \bibinfo{pages}{094003} (\bibinfo{year}{2024}), \eprint{2402.14467}.

\bibitem[{\citenamefont{Wang et~al.}(2021)\citenamefont{Wang, Chen, and
  Liu}}]{Wang:2020bjt}
\bibinfo{author}{\bibfnamefont{F.-L.} \bibnamefont{Wang}},
  \bibinfo{author}{\bibfnamefont{R.}~\bibnamefont{Chen}}, \bibnamefont{and}
  \bibinfo{author}{\bibfnamefont{X.}~\bibnamefont{Liu}},
  \bibinfo{journal}{Phys. Rev. D} \textbf{\bibinfo{volume}{103}},
  \bibinfo{pages}{034014} (\bibinfo{year}{2021}), \eprint{2011.14296}.

\bibitem[{\citenamefont{Ferretti and Santopinto}(2020)}]{Ferretti:2020ewe}
\bibinfo{author}{\bibfnamefont{J.}~\bibnamefont{Ferretti}} \bibnamefont{and}
  \bibinfo{author}{\bibfnamefont{E.}~\bibnamefont{Santopinto}},
  \bibinfo{journal}{JHEP} \textbf{\bibinfo{volume}{04}}, \bibinfo{pages}{119}
  (\bibinfo{year}{2020}), \eprint{2001.01067}.

\bibitem[{\citenamefont{Azizi et~al.}(2022)\citenamefont{Azizi, Sarac, and
  Sundu}}]{Azizi:2021pbh}
\bibinfo{author}{\bibfnamefont{K.}~\bibnamefont{Azizi}},
  \bibinfo{author}{\bibfnamefont{Y.}~\bibnamefont{Sarac}}, \bibnamefont{and}
  \bibinfo{author}{\bibfnamefont{H.}~\bibnamefont{Sundu}},
  \bibinfo{journal}{Eur. Phys. J. C} \textbf{\bibinfo{volume}{82}},
  \bibinfo{pages}{543} (\bibinfo{year}{2022}), \eprint{2112.15543}.

\bibitem[{\citenamefont{Mars\'e-Valera
  et~al.}(2023)\citenamefont{Mars\'e-Valera, Magas, and
  Ramos}}]{Marse-Valera:2022khy}
\bibinfo{author}{\bibfnamefont{J.~A.} \bibnamefont{Mars\'e-Valera}},
  \bibinfo{author}{\bibfnamefont{V.~K.} \bibnamefont{Magas}}, \bibnamefont{and}
  \bibinfo{author}{\bibfnamefont{A.}~\bibnamefont{Ramos}},
  \bibinfo{journal}{Phys. Rev. Lett.} \textbf{\bibinfo{volume}{130}},
  \bibinfo{pages}{091903} (\bibinfo{year}{2023}), \eprint{2210.02792}.

\bibitem[{\citenamefont{Roca et~al.}(2024)\citenamefont{Roca, Song, and
  Oset}}]{Roca:2024nsi}
\bibinfo{author}{\bibfnamefont{L.}~\bibnamefont{Roca}},
  \bibinfo{author}{\bibfnamefont{J.}~\bibnamefont{Song}}, \bibnamefont{and}
  \bibinfo{author}{\bibfnamefont{E.}~\bibnamefont{Oset}},
  \bibinfo{journal}{Phys. Rev. D} \textbf{\bibinfo{volume}{109}},
  \bibinfo{pages}{094005} (\bibinfo{year}{2024}), \eprint{2403.08732}.

\bibitem[{\citenamefont{Song et~al.}(2024)\citenamefont{Song, Duan, Roca, and
  Oset}}]{Song:2024yli}
\bibinfo{author}{\bibfnamefont{J.}~\bibnamefont{Song}},
  \bibinfo{author}{\bibfnamefont{M.-Y.} \bibnamefont{Duan}},
  \bibinfo{author}{\bibfnamefont{L.}~\bibnamefont{Roca}}, \bibnamefont{and}
  \bibinfo{author}{\bibfnamefont{E.}~\bibnamefont{Oset}},
  \bibinfo{journal}{Eur. Phys. J. C} \textbf{\bibinfo{volume}{84}},
  \bibinfo{pages}{1055} (\bibinfo{year}{2024}), \eprint{2406.14895}.

\bibitem[{\citenamefont{Nakano et~al.}(2003)}]{LEPS:2003wug}
\bibinfo{author}{\bibfnamefont{T.}~\bibnamefont{Nakano}} \bibnamefont{et~al.}
  (\bibinfo{collaboration}{LEPS}), \bibinfo{journal}{Phys. Rev. Lett.}
  \textbf{\bibinfo{volume}{91}}, \bibinfo{pages}{012002}
  (\bibinfo{year}{2003}), \eprint{hep-ex/0301020}.

\bibitem[{\citenamefont{Martinez~Torres and
  Oset}(2010{\natexlab{a}})}]{MartinezTorres:2010zzb}
\bibinfo{author}{\bibfnamefont{A.}~\bibnamefont{Martinez~Torres}}
  \bibnamefont{and} \bibinfo{author}{\bibfnamefont{E.}~\bibnamefont{Oset}},
  \bibinfo{journal}{Phys. Rev. Lett.} \textbf{\bibinfo{volume}{105}},
  \bibinfo{pages}{092001} (\bibinfo{year}{2010}{\natexlab{a}}),
  \eprint{1008.4978}.

\bibitem[{\citenamefont{Martinez~Torres and
  Oset}(2010{\natexlab{b}})}]{MartinezTorres:2010xqq}
\bibinfo{author}{\bibfnamefont{A.}~\bibnamefont{Martinez~Torres}}
  \bibnamefont{and} \bibinfo{author}{\bibfnamefont{E.}~\bibnamefont{Oset}},
  \bibinfo{journal}{Phys. Rev. C} \textbf{\bibinfo{volume}{81}},
  \bibinfo{pages}{055202} (\bibinfo{year}{2010}{\natexlab{b}}),
  \eprint{1003.1098}.

\bibitem[{\citenamefont{Navas et~al.}(2024)}]{ParticleDataGroup:2024cfk}
\bibinfo{author}{\bibfnamefont{S.}~\bibnamefont{Navas}} \bibnamefont{et~al.}
  (\bibinfo{collaboration}{Particle Data Group}), \bibinfo{journal}{Phys. Rev.
  D} \textbf{\bibinfo{volume}{110}}, \bibinfo{pages}{030001}
  (\bibinfo{year}{2024}).

\bibitem[{\citenamefont{Kolomeitsev and Lutz}(2004)}]{Kolomeitsev:2003kt}
\bibinfo{author}{\bibfnamefont{E.~E.} \bibnamefont{Kolomeitsev}}
  \bibnamefont{and} \bibinfo{author}{\bibfnamefont{M.~F.~M.}
  \bibnamefont{Lutz}}, \bibinfo{journal}{Phys. Lett. B}
  \textbf{\bibinfo{volume}{585}}, \bibinfo{pages}{243} (\bibinfo{year}{2004}),
  \eprint{nucl-th/0305101}.

\bibitem[{\citenamefont{Xu et~al.}(2016)\citenamefont{Xu, Xie, Chen, and
  Jia}}]{Xu:2015bpl}
\bibinfo{author}{\bibfnamefont{S.-Q.} \bibnamefont{Xu}},
  \bibinfo{author}{\bibfnamefont{J.-J.} \bibnamefont{Xie}},
  \bibinfo{author}{\bibfnamefont{X.-R.} \bibnamefont{Chen}}, \bibnamefont{and}
  \bibinfo{author}{\bibfnamefont{D.-J.} \bibnamefont{Jia}},
  \bibinfo{journal}{Commun. Theor. Phys.} \textbf{\bibinfo{volume}{65}},
  \bibinfo{pages}{53} (\bibinfo{year}{2016}), \eprint{1510.07419}.

\bibitem[{\citenamefont{Wang et~al.}(2008)\citenamefont{Wang, Huang, Zhang, and
  Liu}}]{Wang:2008zzz}
\bibinfo{author}{\bibfnamefont{W.~L.} \bibnamefont{Wang}},
  \bibinfo{author}{\bibfnamefont{F.}~\bibnamefont{Huang}},
  \bibinfo{author}{\bibfnamefont{Z.~Y.} \bibnamefont{Zhang}}, \bibnamefont{and}
  \bibinfo{author}{\bibfnamefont{F.}~\bibnamefont{Liu}}, \bibinfo{journal}{J.
  Phys. G} \textbf{\bibinfo{volume}{35}}, \bibinfo{pages}{085003}
  (\bibinfo{year}{2008}).

\bibitem[{\citenamefont{Gamermann et~al.}(2011)\citenamefont{Gamermann,
  Garcia-Recio, Nieves, and Salcedo}}]{Gamermann:2011mq}
\bibinfo{author}{\bibfnamefont{D.}~\bibnamefont{Gamermann}},
  \bibinfo{author}{\bibfnamefont{C.}~\bibnamefont{Garcia-Recio}},
  \bibinfo{author}{\bibfnamefont{J.}~\bibnamefont{Nieves}}, \bibnamefont{and}
  \bibinfo{author}{\bibfnamefont{L.~L.} \bibnamefont{Salcedo}},
  \bibinfo{journal}{Phys. Rev. D} \textbf{\bibinfo{volume}{84}},
  \bibinfo{pages}{056017} (\bibinfo{year}{2011}), \eprint{1104.2737}.

\bibitem[{\citenamefont{Oset and Ramos}(1998)}]{Oset:1997it}
\bibinfo{author}{\bibfnamefont{E.}~\bibnamefont{Oset}} \bibnamefont{and}
  \bibinfo{author}{\bibfnamefont{A.}~\bibnamefont{Ramos}},
  \bibinfo{journal}{Nucl. Phys. A} \textbf{\bibinfo{volume}{635}},
  \bibinfo{pages}{99} (\bibinfo{year}{1998}), \eprint{nucl-th/9711022}.

\bibitem[{\citenamefont{Hyodo and Jido}(2012)}]{Hyodo:2011ur}
\bibinfo{author}{\bibfnamefont{T.}~\bibnamefont{Hyodo}} \bibnamefont{and}
  \bibinfo{author}{\bibfnamefont{D.}~\bibnamefont{Jido}},
  \bibinfo{journal}{Prog. Part. Nucl. Phys.} \textbf{\bibinfo{volume}{67}},
  \bibinfo{pages}{55} (\bibinfo{year}{2012}), \eprint{1104.4474}.

\bibitem[{\citenamefont{Oller and Meissner}(2001)}]{Oller:2000fj}
\bibinfo{author}{\bibfnamefont{J.~A.} \bibnamefont{Oller}} \bibnamefont{and}
  \bibinfo{author}{\bibfnamefont{U.~G.} \bibnamefont{Meissner}},
  \bibinfo{journal}{Phys. Lett. B} \textbf{\bibinfo{volume}{500}},
  \bibinfo{pages}{263} (\bibinfo{year}{2001}), \eprint{hep-ph/0011146}.

\bibitem[{\citenamefont{Scherer}(2003)}]{Scherer:2002tk}
\bibinfo{author}{\bibfnamefont{S.}~\bibnamefont{Scherer}},
  \bibinfo{journal}{Adv. Nucl. Phys.} \textbf{\bibinfo{volume}{27}},
  \bibinfo{pages}{277} (\bibinfo{year}{2003}), \eprint{hep-ph/0210398}.

\bibitem[{\citenamefont{Gasser et~al.}(1991)\citenamefont{Gasser, Leutwyler,
  and Sainio}}]{Gasser:1990ce}
\bibinfo{author}{\bibfnamefont{J.}~\bibnamefont{Gasser}},
  \bibinfo{author}{\bibfnamefont{H.}~\bibnamefont{Leutwyler}},
  \bibnamefont{and} \bibinfo{author}{\bibfnamefont{M.~E.}
  \bibnamefont{Sainio}}, \bibinfo{journal}{Phys. Lett. B}
  \textbf{\bibinfo{volume}{253}}, \bibinfo{pages}{252} (\bibinfo{year}{1991}).

\bibitem[{\citenamefont{Feijoo et~al.}(2023{\natexlab{b}})\citenamefont{Feijoo,
  Valcarce~Cadenas, and Magas}}]{Feijoo:2023wua}
\bibinfo{author}{\bibfnamefont{A.}~\bibnamefont{Feijoo}},
  \bibinfo{author}{\bibfnamefont{V.}~\bibnamefont{Valcarce~Cadenas}},
  \bibnamefont{and} \bibinfo{author}{\bibfnamefont{V.~K.} \bibnamefont{Magas}},
  \bibinfo{journal}{Phys. Lett. B} \textbf{\bibinfo{volume}{841}},
  \bibinfo{pages}{137927} (\bibinfo{year}{2023}{\natexlab{b}}),
  \bibinfo{note}{[Erratum: Phys.Lett.B 853, 138660 (2024)]},
  \eprint{2303.01323}.

\bibitem[{\citenamefont{Feijoo et~al.}(2019)\citenamefont{Feijoo, Magas, and
  Ramos}}]{Feijoo:2018den}
\bibinfo{author}{\bibfnamefont{A.}~\bibnamefont{Feijoo}},
  \bibinfo{author}{\bibfnamefont{V.}~\bibnamefont{Magas}}, \bibnamefont{and}
  \bibinfo{author}{\bibfnamefont{A.}~\bibnamefont{Ramos}},
  \bibinfo{journal}{Phys. Rev. C} \textbf{\bibinfo{volume}{99}},
  \bibinfo{pages}{035211} (\bibinfo{year}{2019}), \eprint{1810.07600}.

\bibitem[{\citenamefont{Sarti et~al.}(2024)\citenamefont{Sarti, Feijoo,
  Vida\~na, Ramos, Giacosa, Hyodo, and Kamiya}}]{Sarti:2023wlg}
\bibinfo{author}{\bibfnamefont{V.~M.} \bibnamefont{Sarti}},
  \bibinfo{author}{\bibfnamefont{A.}~\bibnamefont{Feijoo}},
  \bibinfo{author}{\bibfnamefont{I.}~\bibnamefont{Vida\~na}},
  \bibinfo{author}{\bibfnamefont{A.}~\bibnamefont{Ramos}},
  \bibinfo{author}{\bibfnamefont{F.}~\bibnamefont{Giacosa}},
  \bibinfo{author}{\bibfnamefont{T.}~\bibnamefont{Hyodo}}, \bibnamefont{and}
  \bibinfo{author}{\bibfnamefont{Y.}~\bibnamefont{Kamiya}},
  \bibinfo{journal}{Phys. Rev. D} \textbf{\bibinfo{volume}{110}},
  \bibinfo{pages}{L011505} (\bibinfo{year}{2024}), \eprint{2309.08756}.

\bibitem[{\citenamefont{Bazzi et~al.}(2011)}]{SIDD}
\bibinfo{author}{\bibfnamefont{M.}~\bibnamefont{Bazzi}} \bibnamefont{et~al.}
  (\bibinfo{collaboration}{SIDDHARTA}), \bibinfo{journal}{Phys. Lett. B}
  \textbf{\bibinfo{volume}{704}}, \bibinfo{pages}{113} (\bibinfo{year}{2011}),
  \eprint{1105.3090}.

\bibitem[{\citenamefont{Acharya et~al.}(2023)}]{ALICE:2023wjz}
\bibinfo{author}{\bibfnamefont{S.}~\bibnamefont{Acharya}} \bibnamefont{et~al.}
  (\bibinfo{collaboration}{ALICE}), \bibinfo{journal}{Phys. Lett. B}
  \textbf{\bibinfo{volume}{845}}, \bibinfo{pages}{138145}
  (\bibinfo{year}{2023}), \eprint{2305.19093}.

\bibitem[{\citenamefont{Borasoy et~al.}(2005)\citenamefont{Borasoy, Nissler,
  and Weise}}]{Borasoy:2005ie}
\bibinfo{author}{\bibfnamefont{B.}~\bibnamefont{Borasoy}},
  \bibinfo{author}{\bibfnamefont{R.}~\bibnamefont{Nissler}}, \bibnamefont{and}
  \bibinfo{author}{\bibfnamefont{W.}~\bibnamefont{Weise}},
  \bibinfo{journal}{Eur. Phys. J. A} \textbf{\bibinfo{volume}{25}},
  \bibinfo{pages}{79} (\bibinfo{year}{2005}), \eprint{hep-ph/0505239}.

\bibitem[{\citenamefont{Ramos et~al.}(2016)\citenamefont{Ramos, Feijoo, and
  Magas}}]{Ramos:2016odk}
\bibinfo{author}{\bibfnamefont{A.}~\bibnamefont{Ramos}},
  \bibinfo{author}{\bibfnamefont{A.}~\bibnamefont{Feijoo}}, \bibnamefont{and}
  \bibinfo{author}{\bibfnamefont{V.~K.} \bibnamefont{Magas}},
  \bibinfo{journal}{Nucl. Phys. A} \textbf{\bibinfo{volume}{954}},
  \bibinfo{pages}{58} (\bibinfo{year}{2016}), \eprint{1605.03767}.

\bibitem[{\citenamefont{Feijoo et~al.}(2021)\citenamefont{Feijoo, Gazda, Magas,
  and Ramos}}]{Feijoo:2021zau}
\bibinfo{author}{\bibfnamefont{A.}~\bibnamefont{Feijoo}},
  \bibinfo{author}{\bibfnamefont{D.}~\bibnamefont{Gazda}},
  \bibinfo{author}{\bibfnamefont{V.}~\bibnamefont{Magas}}, \bibnamefont{and}
  \bibinfo{author}{\bibfnamefont{A.}~\bibnamefont{Ramos}},
  \bibinfo{journal}{Symmetry} \textbf{\bibinfo{volume}{13}},
  \bibinfo{pages}{1434} (\bibinfo{year}{2021}), \eprint{2107.10560}.

\bibitem[{\citenamefont{Ikeno et~al.}(2020)\citenamefont{Ikeno, Toledo, and
  Oset}}]{Ikeno:2020vqv}
\bibinfo{author}{\bibfnamefont{N.}~\bibnamefont{Ikeno}},
  \bibinfo{author}{\bibfnamefont{G.}~\bibnamefont{Toledo}}, \bibnamefont{and}
  \bibinfo{author}{\bibfnamefont{E.}~\bibnamefont{Oset}},
  \bibinfo{journal}{Phys. Rev. D} \textbf{\bibinfo{volume}{101}},
  \bibinfo{pages}{094016} (\bibinfo{year}{2020}), \eprint{2003.07580}.

\bibitem[{\citenamefont{Jia et~al.}(2019)}]{Belle:2019zco}
\bibinfo{author}{\bibfnamefont{S.}~\bibnamefont{Jia}} \bibnamefont{et~al.}
  (\bibinfo{collaboration}{Belle}), \bibinfo{journal}{Phys. Rev. D}
  \textbf{\bibinfo{volume}{100}}, \bibinfo{pages}{032006}
  (\bibinfo{year}{2019}), \eprint{1906.00194}.

\bibitem[{\citenamefont{Mikhasenko}(2023)}]{Misha2023}
\bibinfo{author}{\bibfnamefont{M.}~\bibnamefont{Mikhasenko}},
  \emph{\bibinfo{title}{Exotic hadrons from \mbox{LHCb}}}
  (\bibinfo{year}{2023}), \bibinfo{note}{plenary talk on behalf of LHCb
  collaboration at MESON2023, Krakow},
  \urlprefix\url{https://indico.meson.if.uj.edu.pl/event/3/contributions/370/attachments/212/350/Exotic_Hadrons_MESON2023_v1.pdf}.

\bibitem[{\citenamefont{Roca et~al.}(2015)\citenamefont{Roca, Mai, Oset, and
  Mei\ss{}ner}}]{Roca:2015tea}
\bibinfo{author}{\bibfnamefont{L.}~\bibnamefont{Roca}},
  \bibinfo{author}{\bibfnamefont{M.}~\bibnamefont{Mai}},
  \bibinfo{author}{\bibfnamefont{E.}~\bibnamefont{Oset}}, \bibnamefont{and}
  \bibinfo{author}{\bibfnamefont{U.-G.} \bibnamefont{Mei\ss{}ner}},
  \bibinfo{journal}{Eur. Phys. J. C} \textbf{\bibinfo{volume}{75}},
  \bibinfo{pages}{218} (\bibinfo{year}{2015}), \eprint{1503.02936}.

\bibitem[{\citenamefont{Feijoo et~al.}(2015)\citenamefont{Feijoo, Magas, Ramos,
  and Oset}}]{Feijoo:2015cca}
\bibinfo{author}{\bibfnamefont{A.}~\bibnamefont{Feijoo}},
  \bibinfo{author}{\bibfnamefont{V.~K.} \bibnamefont{Magas}},
  \bibinfo{author}{\bibfnamefont{A.}~\bibnamefont{Ramos}}, \bibnamefont{and}
  \bibinfo{author}{\bibfnamefont{E.}~\bibnamefont{Oset}},
  \bibinfo{journal}{Phys. Rev. D} \textbf{\bibinfo{volume}{92}},
  \bibinfo{pages}{076015} (\bibinfo{year}{2015}), \bibinfo{note}{[Erratum:
  Phys.Rev.D 95, 039905 (2017)]}, \eprint{1507.04640}.

\bibitem[{\citenamefont{Feijoo et~al.}(2024)\citenamefont{Feijoo, Sarti,
  Nieves, Ramos, and Vida\~na}}]{Feijoo:2024qqg}
\bibinfo{author}{\bibfnamefont{A.}~\bibnamefont{Feijoo}},
  \bibinfo{author}{\bibfnamefont{V.~M.} \bibnamefont{Sarti}},
  \bibinfo{author}{\bibfnamefont{J.}~\bibnamefont{Nieves}},
  \bibinfo{author}{\bibfnamefont{A.}~\bibnamefont{Ramos}}, \bibnamefont{and}
  \bibinfo{author}{\bibfnamefont{I.}~\bibnamefont{Vida\~na}}
  (\bibinfo{year}{2024}), \eprint{2411.10245}.

\bibitem[{\citenamefont{Koonin}(1977)}]{Koonin:1977fh}
\bibinfo{author}{\bibfnamefont{S.~E.} \bibnamefont{Koonin}},
  \bibinfo{journal}{Phys. Lett. B} \textbf{\bibinfo{volume}{70}},
  \bibinfo{pages}{43} (\bibinfo{year}{1977}).

\bibitem[{\citenamefont{Pratt}(1986)}]{Pratt:1986cc}
\bibinfo{author}{\bibfnamefont{S.}~\bibnamefont{Pratt}},
  \bibinfo{journal}{Phys. Rev. D} \textbf{\bibinfo{volume}{33}},
  \bibinfo{pages}{1314} (\bibinfo{year}{1986}).

\bibitem[{\citenamefont{Ohnishi et~al.}(2016)\citenamefont{Ohnishi, Morita,
  Miyahara, and Hyodo}}]{Ohnishi:2016elb}
\bibinfo{author}{\bibfnamefont{A.}~\bibnamefont{Ohnishi}},
  \bibinfo{author}{\bibfnamefont{K.}~\bibnamefont{Morita}},
  \bibinfo{author}{\bibfnamefont{K.}~\bibnamefont{Miyahara}}, \bibnamefont{and}
  \bibinfo{author}{\bibfnamefont{T.}~\bibnamefont{Hyodo}},
  \bibinfo{journal}{Nucl. Phys. A} \textbf{\bibinfo{volume}{954}},
  \bibinfo{pages}{294} (\bibinfo{year}{2016}), \eprint{1603.05761}.

\bibitem[{\citenamefont{Haidenbauer}(2019)}]{Haidenbauer:2018jvl}
\bibinfo{author}{\bibfnamefont{J.}~\bibnamefont{Haidenbauer}},
  \bibinfo{journal}{Nucl. Phys. A} \textbf{\bibinfo{volume}{981}},
  \bibinfo{pages}{1} (\bibinfo{year}{2019}).

\bibitem[{\citenamefont{Albaladejo et~al.}(2024)\citenamefont{Albaladejo,
  Feijoo, Nieves, Oset, and Vida\~na}}]{Albaladejo:2024lam}
\bibinfo{author}{\bibfnamefont{M.}~\bibnamefont{Albaladejo}},
  \bibinfo{author}{\bibfnamefont{A.}~\bibnamefont{Feijoo}},
  \bibinfo{author}{\bibfnamefont{J.}~\bibnamefont{Nieves}},
  \bibinfo{author}{\bibfnamefont{E.}~\bibnamefont{Oset}}, \bibnamefont{and}
  \bibinfo{author}{\bibfnamefont{I.}~\bibnamefont{Vida\~na}}
  (\bibinfo{year}{2024}), \eprint{2410.08880}.

\end{thebibliography}

\end{document}